\documentclass[11pt]{article}
\usepackage{amsmath}
\usepackage{multicol}
\usepackage{graphicx}
\usepackage{pstricks}
\usepackage{pst-plot}
\usepackage{floatflt}
\usepackage{epsfig}
\usepackage[french]{babel}
\usepackage[latin1]{inputenc}

\def\bq{\begin{quote}\em\small}
\def\eq{\end{quote}}

\def\be{\begin{equation}}
\def\ee{\end{equation}}

\def\bea{\begin{eqnarray}}
\def\eea{\end{eqnarray}}
\def\beann{\begin{eqnarray*}}
\def\eeann{\end{eqnarray*}}

\newcommand{\ket}[1]{|\kern.3ex #1 \kern.3ex\rangle}
\newcommand{\bra}[1]{\langle\kern.3ex #1 \kern.3ex|}
\newcommand{\braket}[2]{\langle\kern.3ex #1 \kern.3ex|
                        \kern.3ex #2 \kern.3ex\rangle}

\def\chu{ 
\begin{pspicture}(.19,.2)
\psline{-}(0,-.09)(.15,.21)
\pscircle*(.15,.21){.05}
\end{pspicture}
}
\def\chd{ 
\begin{pspicture}(.19,.2)
\psline{-}(0,.24)(.15,-.06)
\pscircle*(.15,-.06){.05}
\end{pspicture}
}
\def\dgu{\mathrm{D}_g\!\!\uparrow}
\def\dgd{\mathrm{D}_{g}\!\downarrow}
\def\deu{\mathrm{D}_e\!\!\uparrow}
\def\ded{\mathrm{D}_{e}\!\downarrow}
\begin{document}
\title{
L'ARGUMENT EINSTEIN PODOLSKY ROSEN (EPR) : PARADOXE, ALTERNATIVE ET
DÉMONSTRATION }
\author{ P.Roussel,
Institut de Physique nucl\'eaire\\
Universit\'e Paris XI, CNRS, IN2P3 \\
F-91406 Orsay Cedex }
\date{}
\maketitle
\begin {abstract}

En physique quantique, l'absence d'observation d'états de superposition ne
devrait être une surprise pour personne, que les objets soient
microscopiques ou macroscopiques. Cela n'en était pas une pour Schrödinger.
L'expérience de pensée ``du chat'' parmi bien d'autres arguments le montre
clairement. On ne doit pas aujourd'hui lui faire dire le contraire.
L'article EPR (A. Einstein, B. Podolsky, N. Rosen, écrit en 1935), qui est à
l'origine de l'intervention de Schrödinger et du développement d'une
problématique qui se prolonge jusqu'aujourd'hui, est analysé. Avec
Schrödinger, Bohr, Wigner et jusqu'à aujourd'hui, la réponse à EPR a été le
recours renouvelé à l'observateur finalement caractérisé par sa conscience. Deux
éléments nouveaux sont apparus : d'une part un développement expérimental et
théorique impressionnant avec l'optique quantique et les manipulations
d'atomes, d'autre part la proposition du concept de {\em décohérence}
susceptible d'expliquer le passage du micro- au macroscopique et par là
d'apporter un élément de compréhension de la {\em mesure} quantique.  On
montre que les questions abordées par EPR restent plus que jamais posées
mais que les expériences aujourd'hui accessibles permettent d'y travailler
concrètement, ramenant dans le domaine de la physique ce qui était évacué
vers l'interprétation ou neutralisé par l'introduction de termes aussi
incontestables qu'inefficaces comme ``non-localité'' ou ``non-séparabilité''.
Einstein ne pourrait que s'en réjouir.

\end {abstract}

\tableofcontents
\section {Introduction.}

La présentation des réalisations contemporaines remarquables de l'optique
quantique associée à la physique atomique s'accompagne fréquemment de
l'invocation du ``chat de Schrödinger''. L'expérience de pensée du chat 
y est utilisée pour illustrer le problème de la mesure en mécanique quantique 
(MQ) et le rattacher
d'une part à ce qui est appelé la transition du micro- au macroscopique et
d'autre part à la ``décohérence'', un concept introduit pour rendre compte,
au moins partiellement, de façon plus scientifique justement, de ce qui se
passe -on ne peut pas être plus précis pour le moment- au moment de la
mesure, de la réduction du paquet d'onde.

Il se trouve que l'expérience de pensée du chat -au demeurant un élément
mineur de l'argumentation de Schrödinger- est utilisée à contresens
(peut-être depuis longtemps) mais que la mise en perspective de tout ce
débat a un point de départ -au moins un point de passage fort et obligé -
l'article EPR publié peu de temps avant celui de Schrödinger et dont les
conséquences et les enjeux se poursuivent jusqu'à aujourd'hui. On se doit de
dégager ce qu'il faut appeler l'alternative EPR et au-delà la démonstration
EPR, ce qui est bien autre chose qu'un paradoxe.

On examinera si les développements expérimentaux et théoriques récents et
l'introduction du concept de décohérence sont susceptibles de répondre aux
interrogations de cette alternative et d'éclairer le processus de mesure
quantique autrement que par le recours à la conscience de l'observateur
 utilisé jusqu'ici.

\section{Le chat de Schrödinger, un agent retourné.}

\subsection{Une affirmation répétée concernant les états de superposition.}

Dans des articles de vulgarisation ou dans des articles plus directement
scientifiques
on a vu Schrödinger maintes fois invoqué au travers du ``paradoxe'' de son
chat, pour s'étonner de l'absence d'observation d'états de superposition
dans le monde macroscopique, au contraire dit-on du monde microscopique, et
pour s'interroger alors sur la transition d'un monde à l'autre.
L'appellation ``état-chat'' est même devenue courante pour des états de
superposition concernant des objets mésoscopiques.  On se rappellera les
multiples ``représentations'' graphiques, construites pour choquer,
de ce que justement on ne voit pas : un chat à moitié mort à moitié vivant.
Mais examinons les textes et commençons avec W.H. Zurek (1991) \cite{Zure91} :

\bq
Given almost any initial condition the universe described by
\textbar$\Psi$\textgreater $~$ evolves
into a state that simultaneously contains many alternatives never seen to
coexist in our world.\ldots 

Thus at the root of our unease with quantum mechanics is the clash between
the principle of superposition -the consequence of the linearity of
[the Schrödinger] equation- and the everyday classical reality in which this
principle appears to be violated.\ldots 

Delineating the border between the quantum realm ruled by the Schrödinger
equation and the classical realm ruled by the Newton's laws is one of the
unresolved problem of physics {\em (cette dernière phrase dans la légende de la
figure 1, p. 37).}
\eq

Citons maintenant plusieurs membres d'une équipe de recherche de l'ENS Paris
dont l'apport dans le domaine de l'optique quantique et la manipulation
d'atomes est essentielle. D'abord, M. Brune et al (1996) \cite{Brun96} :

\bq
The transition between the microscopic and macroscopic worlds is a
fundamental issue in quantum measurement theory\ldots . [A quantum
superposition state of the ``meter + atom'' system] is however never
observed. Schrödinger has vividly illustrated this problem by replacing the
meter by a ``cat'' and considering the dramatic superposition of dead and
alive animal states. Although such a striking image can only be a metaphor,
quantum superpositions involving ``meter states'' are often called
``Schrödinger cats''.
\eq

Ensuite, J.M. Raimond et al (1997) \cite{Raim97} :
\bq
The absence of macroscopic quantum superpositions is a central issue  in
our understanding of quantum measurement theory\ldots 

Schrödinger has vividly illustrated this problem by replacing the
meter by a cat whose life is dependent upon the fate of a radioactive atom.
This situation leads to the paradoxical superposition of ``dead'' and
``alive'' cat states.
\eq

Enfin S. Haroche et al(1997) \cite{Haro97} :
\bq
Un chat [peut-il être] {\em simultanément} vivant et mort? Ces
questions burlesques ont été posées\ldots  par Albert Einstein et Erwin
Schrödinger qui cherchaient à appliquer directement à notre monde les
concepts de la mécanique quantique\ldots
\eq
et en légende de la photo ``représentant'' un chat à  moitié mort, à moitié
vivant :
\bq
{\bf Ce montage photographique représente-t-il une situation réelle?}

Les principes de la physique quantique n'interdisent pas à un objet
macroscopique d'être dans deux états à la fois. De telles situations ne sont
pourtant jamais observées. C'est le paradoxe du chat de Schrödinger, du
nom du physicien autrichien qui, en 1935, imagina un scénario où un chat
devrait être à la fois vivant et mort.
\eq

Tout récemment encore \cite{Raim01} avec un des auteurs précédents :

\bq
The superposition principle is at the heart of the most intriguing features
of the microscopic world\ldots  \ldots It is impossible to get a classical
intuitive representation of these superpositions. Their oddity becomes
evident when one transpose them to the macroscopic scale, as in the famous
``Schrödinger cat'' metaphor (Schrödinger, 1935), describing a cat suspended
between life and death.
\eq

Et revenons à W.H. Zurek, en 1997 \cite{Zure97}:
\bq
In other words, the apparatus itself [if quantum mechanics is universal] is
described by fuzzy quantum wave function, rather than by a definite
classical state. The famous Schrödinger cat attains its suspension between
life and death in this manner.
\eq

J.A. Wheeler est un des physiciens qui sont beaucoup intervenus dans les
débats sur les fondements de la MQ. Il s'exprime ici très récemment encore
avec M. Tegmark (2001) \cite{Whee01} :
\bq
Schrödinger fit remarquer que si des objets microscopiques tels que des
atomes, peuvent se trouver dans des superpositions, il pourrait en être de
même pour des objets macroscopiques, puisque ceux-cis sont constitués
d'atomes\ldots Comme l'atome radioactif va se trouver dans une superposition
d'un état désintégré et d'un état non désintégré, il produit un chat à la
fois vivant et mort, en superposition des deux.
\eq

Terminons avec le bulletin de la société américaine de physique
 \cite{Baps02} qui marque, précisément en ce mois de mars (2002)
l'anniversaire de la parution de l'article de Schrödinger :
\bq
Schrödinger used the analogy to demonstrate the limitations of quantum
mechanics : quantum particles such as atoms can be in two or more different
quantum states at the same time, but surely, he argued, a classical object
made of a large number of atoms, such as a cat, should not be in two
different states.
\eq

Deux questions se posent alors : que dit effectivement la mécanique
quantique (éventuellement, que disait-elle) de ces états de superposition?
Que disait Schrödinger avec le ``paradoxe du chat''?

\subsection{Une réalité en contradiction avec cette
affirmation.}
\subsubsection{Ce que dit la mécanique quantique (et ce qu'elle disait).}

Rappelons d'abord les postulats \cite{Cohe77} et complétons par le schéma
illustrant le  processus (!) de la mesure.

\vspace{1cm}

\fbox{\parbox{15cm}{{\em 1er Postulat : }A un instant $t_o$ fixé,
l'état d'un système physique est défini par la donnée d'un ket
$\ket{\psi(t_o)}$
appartenant à l'espace des états 
$ \mathcal{E}$. }}

\fbox{\parbox{15cm}{{\em 2ème Postulat : } Toute grandeur physique mesurable
$ \mathcal{A}$ est décrite par un opérateur {\em A} agissant dans
$\mathcal{E}$ ; cet opérateur est une observable.}}

\fbox{\parbox{15cm}{{\em 3ème Postulat : } La mesure d'une grandeur physique
$ \mathcal{A}$ ne peut donner comme résultat qu'une des valeurs propres de
l'observable {\em A} correspondante.
}}

\fbox{\parbox{15cm}{{\em 4ème Postulat (cas d'un spectre discret non dégénéré)
: } Lorsqu'on mesure la grandeur physique $ \mathcal{A}$ sur un système
dans l'état $\psi$ normé, la probabilité $ \mathcal{P}${\em (a$_n$)}
d'obtenir comme résultat la valeur propre non dégénérée {\em a$_n$} de
l'observable {\em A} correspondante est :

$ \mathcal{P}${\em (a$_n$)}=\textbar$\braket{u_n}{\psi}$\textbar$^2$

où $\ket{u_n}$ est le vecteur propre normé de {\em A} associé à la valeur
propre a$_n$.
}}

\fbox{\parbox{15cm}{{\em 5ème Postulat :}
Si la mesure de la grandeur physique $ \mathcal{A}$ sur le système dans
l'état $\psi$ donne le résultat a$_n$, l'état du système immédiatement après
la mesure est la projection normée,
$\frac{P_n\psi}{\sqrt{\bra{\psi}P_n\ket{\psi}}}~$
de $\psi$ sur le sous-espace propre associé à a$_n$.
}}

\newpage
\Large
\hspace{4cm}\fbox{\parbox{5cm}{
\centerline{\bf Mesure donnant}

\centerline{\bf le résultat{\em a$_n$}}}}

\begin{pspicture}(16,5)

\psline[linewidth=.06]{c->}(6.6,5)(6.6,4)
\psline[linewidth=.06]{c-}(6.5,4)(6.5,0)
\psline[linewidth=.06]{c-}(6.7,4)(6.7,0)
\psline[linewidth=.05]{c->}(0,0)(13.5,0)
\psline[linewidth=.05]{c-}(1,.3)(1,-.3)
\psline[linewidth=.05]{c-}(11.75,.3)(11.75,-.3)
\rput{0}(13.7,0){{\em t}}
\rput{0}(11.75,-.5){{\em t$_1$}}
\rput{0}(6.6,-.5){{\em t$_0$}}
\rput{0}(1,-0.5){0}
\rput[bl]{0}(0.,1){$\ket{\psi(0)}$}
\rput[bl]{0}(4.8,1){$\ket{\psi(t_o)}$}
\rput[bl]{0}(6.9,3){$\ket{u_n}$}
\rput[bl]{0}(11.,3){$\ket{\psi'(t_1)}$}

\multips{0}(1.8,1)(.8,0){3}{
\pscurve[linewidth=.05](0,.15)(0.2,0)(.4,.15)(.6,.3)(.8,0.15)
}
\psline[linewidth=.05]{c->}(4.2,1.15)(4.5,1.15)

\multips{0}(8.,3)(.8,0){3}{
\pscurve[linewidth=.05](0,.15)(0.2,0)(.4,.15)(.6,.3)(.8,0.15)
}
\psline[linewidth=.05]{c->}(10.4,3.15)(10.7,3.15)
\end{pspicture}

\vspace{7.mm}

\normalsize
\noindent FIGURE 1

{\bf Lors d'une mesure à l'instant {\em t$_0$} de l'observable {\em A}
donnant le résultat {\em a$_n$}, le vecteur d'état du système subit une
brusque modification, et devient $\ket{u_n}$. Il évolue ensuite à partir de
ce nouvel état initial. }

\vspace{9mm}

Le premier postulat est clair, à des questions de norme près : dans un
espace vectoriel, le principe de superposition s'applique (si $\ket{a}$
et$\ket{b}$ sont des états de $\mathcal{E}$, alors ($\ket{a}$ + $\ket{b}$)
l'est aussi) et rien n'indique une limitation concernant la taille du
système. On notera cependant ici le caractère abstrait de ce qu'il faut bien
appeler une représentation du système physique : son état {\em est défini
par la donnée} (d'un ket) ; il n'{\em est} pas une onde ou une fonction
d'onde. On reviendra plus loin avec Schrödinger sur cette question du statut
de $\psi$.

On n'observe ni un ket ni un espace vectoriel mais on effectue des mesures
sur un système physique. Le deuxième postulat dit comment associer 
mesures physiques et espace vectoriel, et le troisième dit très précisément
ce que peut être le résultat d'une mesure : s'il existe une variable, une
observable, à deux valeurs (chat vivant/chat mort), le résultat ne peut être
que l'une de ces valeurs, chacune avec une probabilité évaluée selon le
quatrième postulat. 

 Avec la mécanique quantique, les rapports du réel à l'abstrait ne sont pas
tellement clairs, on verra toute l'attention que Schrödinger a portée à cette
question, mais tout de même, ce ne sont pas les observables qui sont
soumises au principe de superposition, ce sont les ``vecteurs d'état''. Avec
le cinquième postulat on revient au système physique par l'intermédiaire de
son vecteur d'état : le chat est ou bien mort ou bien vivant et il le reste.
Au total, on voit mal qu'on puisse s'étonner de ne pas ``observer'' de chat
à moitié mort à moitié vivant. On pourrait (?) s'étonner de ne pas être
capable de préparer un chat dans un état de superposition, si cela est le
cas, mais l'observation ne donnera jamais que l'un ou l'autre. C'est ce qui
est attendu et c'est ce qui est observé.

Avant d'examiner l'intervention de Schrödinger lui-même, il est utile
d'insister dans cette présentation de la mécanique quantique (celle
d'aujourd'hui) sur la description du processus de la mesure (mais est-ce un
processus? c'est aussi une question).  Depuis la naissance de la mécanique
quantique, c'est une question au c\oe ur 
des discussions c'est aussi un élément qui peut justifier cette
interrogation (évoquée par exemple par Zurek) concernant le passage du
microscopique au macroscopique.

La figure 1, ci-dessus en haut de la page 6 (reproduite de la
référence \cite{Cohe77} page 221), présente parfaitement la question. Le
vecteur d'état (la fonction d'onde) évolue de façon déterministe (équation
de Schrödinger) depuis la préparation initiale en {\em t} = 0 jusqu'à {\bf la
mesure} en {\em t = t$_o$}. Il subit alors un changement brusque probabiliste
vers l'un des vecteurs propres $\ket{u_n}$ de A, celui qui est associé à la
valeur propre a$_n$ trouvée. Il reprend ensuite une évolution déterministe
depuis {\em t = t$_o$} jusqu'à (par exemple) {\em t = t$_1$} où une nouvelle
mesure est éventuellement pratiquée etc\ldots

Ce que la mécanique quantique ne définit justement pas clairement, ce sont
les conditions qui font que l'une ou l'autre des évolutions
(déterministe/probabiliste) doit être choisie. {\bf Une} réponse, c'est que
la ``mesure'', avec l'occurrence de l'aspect probabiliste, fait nécessairement
intervenir un instrument macroscopique, d'où la remarque de W. Zurek
ci-dessus sur le passage du microscopique au macroscopique.

On doit remarquer que cette imprécision de la mécanique quantique n'a, il
est vrai, que rarement ou jamais de conséquence pratique : dans tous les cas
concrets on {\bf sait} appliquer les postulats.

Cette même figure 1 permet aussi d'introduire une distinction essentielle en
mécanique quantique entre ``état pur'' ou vecteur d'état et mélange
statistique\ldots de vecteurs d'état. Le premier est un système physique
``complètement connu'' ; comme ensemble, il est constitué d'éléments
identiques. Par exemple, $\ket{\psi(0)}$, $\ket{\psi(t_o)}$, $\ket{u_n}$ et
$\ket{\psi'(t_1)}$ sont des états purs. Le mélange statistique lui est
incomplètement connu ; comme ensemble, il est constitué d'éléments
différents, chacun avec une probabilité donnée. La mesure sur un ensemble
constitué de kets identiques $\ket{\psi(t_o)}$ fait passer d'un cas pur à un
mélange, celui des cas purs $\ket{u_n}$, chacun avec la probabilité
correspondante donnée par le quatrième postulat (revoir la figure 1, page 6
 ; voir plus loin la citation de London et Bauer page
20).

Avec le principe de superposition, on va dire que

$\ket{a}$, $\ket{b}$, 1/$\sqrt{2}$($\ket{a}$+$\ket{b}$), sont des cas purs. Ce
dernier est bien à distinguer du mélange en proportions égales de $\ket{a}$
et de $\ket{b}$.

\subsubsection{Ce que disait Schrödinger avec le ``paradoxe du chat''.}

On est surpris de ne trouver aucun article sur le ``paradoxe du chat'', ni
sur ``le chat'', mais un long exposé en trois livraisons \cite{Schr35}, avec
en tout 15 chapitres, et intitulés :

\begin{center}
LA SITUATION ACTUELLE DANS LA MÉCANIQUE QUANTIQUE. (I) (II) (III)
\end{center}

(I)\bq 1) La physique des modèles.

2) La statistique des variables des modèles en MQ.

3) Exemples de prédictions de probabilités.

4) Peut-on fonder la théorie sur des ensembles idéaux?

5) Les variables sont-elles réellement brouillées? [verwaschen] [blurred]

\eq

(II)\bq

6) La volte-face délibérée du point de vue épistémologique.

7) La fonction $\Psi$ comme catalogue des valeurs possibles.

8)  Théorie de la mesure (I).

9) La fonction $\Psi$ comme description d'état.

10) Théorie de la mesure (II).
\eq

(III)\bq

11) Résolution de l'enchevêtrement/intrication [Verschränkung]
[entanglement]. Résultat dépendant de l'intention de l'expérimentateur.

12) Un exemple.

13) Continuation de l'exemple : toutes les mesures sont sans équivoque
enchevêtrées.

14) La dépendance temporelle de l'enchevêtrement. Considération du rôle spécial
du temps.

15) Loi naturelle ou moyen de calcul.

\eq

C'est dans le chapitre 5 (Les variables sont-elles réellement brouillées?)
que le fameux chat apparaît. Auparavant, Schrödinger a présenté la physique
classique et l'usage qu'elle fait des modèles. Il a présenté les
spécificités de la mécanique quantique et en particulier le rôle original
qu'y jouent les probabilités. Après avoir montré que la statistique ne peut
pas résulter d'une dispersion déjà présente dans les échantillons, il
s'interroge dans ce chapitre 5 sur la possibilité qu'aurait $\psi$ de se
présenter comme un modèle de la réalité où les variables seraient réellement
``brouillées'', comme un nuage d'électricité autour d'un atome.

Un noyau radioactif est alors utilisé de deux façons -la deuxième pour le
chat. Dans une première expérience de pensée, la source est entourée dans
tout l'espace par un écran luminescent :

\bq The emerging particle is described, if one wants to explain intuitively,
as a spherical wave that continuously emanates in all directions from the
nucleus and that impinges continuously on a surrounding luminescent screen
over its full expanse.
{\bf
\ldots The screen however does not show a more or less constant uniform surface
glow, but rather lights up at one instant at one spot\ldots}
\eq

Dans le même esprit, il remplace l'écran par un détecteur à gaz :

\bq

If \ldots one uses perhaps a gas that is ionised by the $\alpha$-particules,
one finds the ion pairs arranged along rectilinear columns, that project
backwards on to the bit of radioactive matter from which the
$\alpha$-radiation comes.
\eq

Chacun de ces exemples donne déjà la réponse : les variables ne sont pas
brouillées car à chaque fois, on trouve la variable avec {\bf l'une de ses
valeurs possibles}. Pour Schrödinger, et ce n'est pas surprenant, la
mécanique quantique est la même que pour nous aujourd'hui.

Enfin arrive l'expérience du chat, un tiers du chapitre, 1\% de l'article!

\bq 
On can even set up quite ridiculous cases. A cat is penned up in a steel
chamber, along with the following diabolical device (which must be secured
against direct interference by the cat) : in a Geiger counter there is a
tiny bit of radioactive substance, {\em so} small, that {\em perhaps} in the
course of one hour one of the atoms decays, but also, with equal
probability, perhaps none ; if it happens, the counter tube discharges and
through a relay releases a hammer which chatters a small flask of hydrocyanic
acid. If one has left this entire system to itself for an hour, one would
say that the cat still lives {\em if} meanwhile no atom has decayed. The
first atomic decay would have poisoned it.
The $\psi$-function of the entire system would express this by
having in it the living and the dead cat (pardon the expression) mixed or
smeared out in equal parts.

{\bf It is typical of these cases that an indeterminacy originally restricted to
the atomic domain becomes transferred into macroscopic indeterminacy, which
is {\em resolved} by direct observation. 
That prevents us from so
naively accepting as valid a ``blurred model'' for representing reality.}
In itself it would not embody anything unclear or contradictory. There is a
difference between a shaky or out of focus photograph and a snapshot of clouds
and fog banks {\em [fin du chapitre 5].}\eq

Pas plus qu'avec l'expression contemporaine de la mécanique quantique, il
n'y a pas de surprise chez Schrödinger de la non-observation d'un chat à
moitié vivant et mort, pas de paradoxe non plus. On n'attend pas plus
d'observation brouillée de chat vivant et de chat mort que, par exemple,
d'observation brouillée de la localisation d'une particule décrite par une
fonction d'onde étendue.

Il est utile de montrer que le problème n'est pas là puisqu'on répète le
contraire, mais deux questions se posent alors : 

1) pourquoi cette erreur
répétée? -et on proposera plus tard une réponse possible- 

2) quelle est la véritable raison d'être de ce long article de Schrödinger?
Pour  répondre à cette question, on va 
examiner plus en détail le contenu de cet article et
rappeler dans quel contexte il arrive.

On
est conduit par Schrödinger lui-même à un premier retour en arrière
(quelques mois) vers ce qui sera notre base de départ. Dans le  chapitre 12
de son
article, ``Un exemple'',  Schrödinger  indique en effet avec humour, dans
une note en bas de page:

\bq C'est la publication de ce travail \cite{EPR} {\em [ce qui est devenu pour
nous l'article EPR, Einstein, Podolsky, Rosen, tellement cité (peut-être
beaucoup moins lu!)]} qui a motivé la présente -est-ce que je dirai conférence
ou confession générale-? \eq

\section{La cause de tout : l'article EPR, Einstein, Podolsky, Rosen.}
\subsection{Une mise en perspective.}

Avant d'examiner le contenu de l'article EPR \cite{EPR} et les réponses de
Schrödinger, on va remonter encore un peu plus en arrière afin d'éclairer
par deux citations les points de vue d'une part d'Einstein, et d'autre part
ce qu'on appelle le point de vue ou l'école de Copenhague. Einstein d'abord,
en 1926, dans une lettre à Max Born rapportée par Michel Paty \cite{Paty85}
page 335 : 

\bq
La mécanique quantique force le respect. Mais une voix intérieure me dit que
ce n'est pas encore le nec plus ultra. La théorie nous apporte beaucoup de
choses, mais elle nous rapproche à peine du secret du Vieux. De toute façon,
je suis convaincu que lui, au moins, ne joue pas aux dés.
\eq

Sur le fond, trois éléments sont présentés, qu'on retrouvera (plus loin dans
cet article) en 1949 et dont on pourra évaluer les évolutions. Il y a le
respect pour les succès de la MQ, l'insatisfaction globale, et celle
particulière concernant l'aspect probabiliste des résultats.

Cela s'oppose à la déclaration on ne peut plus dogmatique de M. Born et W.
Heisenberg au Conseil Solvay à Bruxelles en 1927, rapportée également par M.
Paty \cite{Paty85}, pages 328-329 :
\bq Nous tenons la mécanique des quanta pour une théorie complète dont les
hypothèses fondamentales physiques et mathématiques ne sont plus susceptibles
de modifications.
\eq

En 1935, Einstein a gardé  le même
état d'esprit qu'en 1926, avec  l'article EPR dont le titre même montre
l'esprit interrogateur :
\bq
Can quantum-mechanical description of physical reality be considered
complete?
\eq
On examine maintenant le contenu de cet article.

\subsection{Le c\oe ur fonctionnel de l'article EPR : la réalité d'un paradoxe.}
Ce qu'on peut appeler le c\oe ur fonctionnel de l'article EPR est important
pour la MQ, pour sa compréhension comme pour son histoire. D'un côté, ce
c\oe ur fonctionnel explicite un processus particulier de mesure, il met en
évidence ce qui plus tard sera traduit par l'introduction de -un mot
mouveau- la non-localité, il est à l'origine de la notion de ``paires EPR'';
d'autre part, il a servi à Einstein depuis 1933 (selon ce qu'en a rapporté L.
Rosenfeld cité dans \cite{Paty85} pages 341 et 342) jusqu'en 1949
(on y reviendra plus loin) à mettre en question la MQ. En voici le schéma :

\begin{multicols}{2}
\resizebox{8.cm}{7.cm}{\LARGE{
\hspace{-1cm}\begin{tabular}{ccc}
$~$\hspace{5cm}$~$&$~$\hspace{3.5cm}$~$&$~$\hspace{5cm}$~$\\
{\Huge(I)} &&\Huge {(II)}\\
\begin{pspicture}(0,0)(5,3.5)
\psline[linewidth=.1]{->}(2.5,3.5)(5,0)
\end{pspicture}
&&
\begin{pspicture}(0,0)(5,3.5)
\psline[linewidth=.1]{->}(2.5,3.5)(0,0)
\end{pspicture}
\\
&&\\
&&\multicolumn{1}{l}{t=0}\\
&
\begin{pspicture}(0,0)(3.5,3.5)
\psline[linewidth=.05]{-}(0.5,-1)(0,-0.5)(0,4)(0.5,4.5)
\psline[linewidth=.05]{-}(3,-1)(3.5,-0.5)(3.5,4)(3,4.5)
\end{pspicture}
&\\
&&\multicolumn{1}{l}{t=T}\\
\begin{pspicture}(0,0)(5,3.5)
\psline[linewidth=.1]{->}(5,3.5)(2.5,0)
\end{pspicture}
&
&
\begin{pspicture}(0,0)(5,3.5)
\psline[linewidth=.1]{->}(0,3.5)(2.5,0)
\end{pspicture}
\\
{\Huge(II)} &&\Huge {(I)}\\
&&\multicolumn{1}{l}{A }\\
&&\multicolumn{1}{l}{ a$_k$}\\
\multicolumn{1}{l}{$\psi_k$(x$_2$)}&
\begin{pspicture}(0,0)(3.5,.5)
\psline[linewidth=.1,linecolor=red]{->}(3.5,.15)(-3.5,.15)
\end{pspicture}
&\multicolumn{1}{l}{$u_k$(x$_1$)}\\
\multicolumn{1}{l}{(par exemple, valeur p$_k$ de P)}&&\\
&&\multicolumn{1}{c}{ou}\\
&&\multicolumn{1}{r}{B}\\
&&  \multicolumn{1}{r}{b$_r$}\\
\multicolumn{1}{r}{$\phi_r$(x$_2$)}&
\begin{pspicture}(0,0)(3.5,.5)
\psline[linewidth=.1,linecolor=red]{->}(7,0.15)(0,0.15)
\end{pspicture}
&\multicolumn{1}{r}{$\nu_r$(x$_1$)}\\
\multicolumn{1}{r}{(par exemple, valeur q$_r$ de Q)}&&\\
\end{tabular}
}
}

$~$
\vspace{-3mm}

\bq \ldots  let us suppose that we have two systems, I and II, which we permit
to interact from the time {\em t=0} to {\em t}={\em T}, after which time we
suppose that there is no longer any interaction between the two parts.

We suppose further that the states of the two systems before {\em t=0} were
known [des états purs]. We can then calculate with the help of the
Schrödinger's equation the state of the combined system I+II at any
subsequent time ; in particular, for any {\em t}\textgreater{\em T}. Let us
designate the corresponding wave function by $\Psi$.
\eq
\bq We cannot however, calculate the state in which either one of the two
systems is left after the interaction. This, according to quantum mechanics,
can be done only with the help of further measurements, by the process known
as the {\em reduction of the wave packet.} 
\eq
\end{multicols}

\bq
Let us consider the essentials of
this process.\eq

Précisons maintenant que ce qui suit repose entièrement sur le type de
fonction d'onde du système composé qui, après interaction, n'est {\bf plus}
factorisable :

$$ \Psi (x_1,x_2) \neq \phi(x_1).\phi(x_2). $$

\bq Suppose now that the quantity A [some physical quantity pertaining to
system I] is measured and it is found that it has the value a$_k$. It is then
conclude that after the measurement the first system is left in the state
given by  the wave function $u_k$(x$_1$), and that the second system is left
in the state given by the wave function $\psi_k$(x$_2$). This is the process
of reduction of the wave packet.

The  set of functions $u_n$(x$_1$) is determined by the choice of the
physical quantity A.  If, instead of this, we had chosen another quantity,
say B,\ldots 

\ldots and [if now] B is measured and is found to have the value b$_r$ we
concluded that after the measurement the first system is left in the state
given by $\nu_r$(x$_1$) and the second system is left in the state given by
$\phi_r$(x$_2$).

We see therefore that, as a consequence of two different measurements
performed upon the first system, the second system may be left in states
with two different wave functions.\eq

Avant de décrire la structure logique de l'article EPR, il est intéressant
de mettre le paragraphe ci-dessus en rapport avec ce que dira Einstein, 14
ans plus tard, dans sa réponse à Marguenau \cite{Eins49}, page 681. Il dit alors
rapporter le point de vue de Niels Bohr (on verra un peu plus loin qu'en
1935, Bohr ne dit pas la même chose) mais dans ses propres termes (ceux
d'Einstein).

\bq If the partial systems A and B form a total system which is described by
its $\Psi$-function $\Psi$\/(AB), there is no reason why any mutually
independent existence (state of reality) should be ascribed to the partial
systems A and B viewed separately, {\em not even if the partial systems are
spatially separated from each other at the particular time under
consideration.} The assertion that, in this latter case, the real situation
of B could not be (directly) influenced by any measurement taken on A is
therefore, within the framework of quantum theory, unfounded and (as the
paradox shows) unacceptable. \eq

Les deux préoccupations (1935 et 1949) coïncident, on peut voir là la
naissance de ``l'inséparabilité quantique'' par sa mise en évidence mais
aussi en question par Einstein. Mais revenons en 1935.

\newpage

\subsection{La 
structure
logique de l'article EPR : alternative et
finalement  démonstration.}

\begin{multicols}{2}
\fbox{\begin{minipage}{7.cm}
\hspace{.5cm}\begin{tabular}{|c|c|c|}
\hline
Réalité && image scientifique\\
&$\not=$&\\
objective&&de cette réalité\\
\hline
\end{tabular}
\\
{\scriptsize Deux questions pour juger une théorie :}\\
La théorie est-elle correcte?\\
Donne-t-elle une description complète?\\
Pour la MQ, deux variables dont les opérateurs ne commutent pas ne peuvent
avoir toutes les deux simultanément une réalité physique.\\
\fbox{
L'expérience des deux systèmes }
$\Longrightarrow$ Une contradiction qui tourne à une alternative dont les deux
termes se rejoignent :

1) L'origine de la détermination des valeurs des variables est dans le passé
commun des deux systèmes et la MQ n'est (simplement) pas complète.

2) On prend en compte la non-simultanéité des déterminations\ldots 

\fbox {C'est à dire la complémentarité de Bohr}

Cela fait dépendre pour (II) la réalité de P ou de Q d'une mesure sur (I)
avec lequel il n'intéragit pas.

{\large
\begin{center}
\begin{tabular}{|c|}
\hline
No reasonable definition of reality\\
could be expected to permit this.\\
\hline
\end{tabular}
\end{center}
}

La mécanique quantique est encore incomplète puisqu'elle introduit une
dépendance qu'elle nie au départ

\end{minipage}}

Avant d'utiliser l'expérience de pensée aux deux systèmes vue précédemment,
EPR posent ce qu'on pourrait appeler leur pilier épistémologique :

\bq Any serious consideration of a physical theory must take into account
the distinction between the objective reality, which is independent of any
theory, and the physical concepts with which the theory operates. These
concepts are intended to correspond with the objective reality, and by means
of these concepts we picture this reality to ourselves. \eq

Et comme on le verra, c'est sur ce pilier que porteront d'abord les réponses.
EPR précisent ensuite comment la théorie doit être correcte et complète. Ils
utilisent alors l'expérience des deux systèmes et appliquent le résultat
décrit précédemment au cas où les opérateurs correspondant aux deux
variables A et B ne commutent pas, par exemple $x$ et $p_x$. Deux
expériences, dont il faut qu'elles soient répétées depuis la préparation
d'un état initial identique, et donc ``complètement connu'', sont conduites l'une
pour la mesure de A, l'autre pour celle de B (et les deux appareils utilisés
seront différents).
\end{multicols}

Des résultats de ces mesures sur I, on {\bf déduit} les valeurs pour II de
deux variables dont les opérateurs ne commutent pas, en contradiction  en
quelque sorte avec l'affirmation de la mécanique quantique que cela n'est
pas possible. Mais cette contradiction mute immédiatement en une alternative
avec ses deux termes :

i) On prend au sérieux et rigoureusement l'affirmation que les deux systèmes
I et II n'intéragissent plus après t=T, et alors le caractère différé des
mesures de A et B n'a pas d'importance, car les valeurs p$_k$ de P et q$_r$
de Q n'ont pu être déterminées que dans le passé commun de I et de II, avant
t=T. Elles restent ensuite {\em simultanément} déterminées, fixées.

ii) On prend au contraire en compte la non-simultanéité des expériences qui
déterminent, ou bien p$_k$, ou bien q$_r$, et anticipant l'intervention de
Bohr, on invoque cette fois la {\em complémentarité}, mais
alors : \bq
This makes the reality of P or Q [sur le système II] depend upon the process
of measurement carried out on the first system which does not disturb the
second system in any way. No reasonable definition of reality could be
expected to permit this.
\eq

Dans les deux cas, la mécanique quantique est incomplète ou inachevée voire
incohérente : l'alternative devient démonstration. Avec i), la MQ est
incomplète et même incohérente puisque la détermination simultanée des deux
variables P et Q n'est pas prévue et même niée. Avec ii) elle met en
évidence une dépendance qu'elle nie au départ. On peut penser que la
rédaction de 1935 privilégie l'élément i) de l'alternative, il n'en sera
plus de même en 1949. Les progrès ultérieurs de la physique ont renforcé ce
mouvement, on y reviendra bien sûr plus loin.

\subsection{Une première réponse qui manque son but : celle de Bohr.}

Avant la publication de Schrödinger qui nous concerne, une première réponse
est publiée rapidement par Niels Bohr, dans la même revue et avec le même
titre \cite{Bohr35}. Son intérêt est pourtant limité car son argumentation
est fondée presque exclusivement sur la complémentarité :

\bq
\ldots a viewpoint termed ``complementarity'' is explained from which
quantum mechanical description of physical phenomena would seem to fulfill,
within its scope, all rational demands of completeness {\em (\cite{Bohr35}
résumé, page 696)}.
\eq

Mais si la complémentarité permet de répondre au terme i) de l'alternative
de l'argument EPR, elle déclenche par contre le recours au terme ii) et
ne permet pas du tout d'y répondre.

On ne sera pas surpris de la déclaration plutôt dogmatique de Bohr :

\bq
Such an argumentation [celle d'EPR] however, would hardly seem suited to
affect the soundness of quantum-mechanical description, which is based on a
coherent mathematical formalism covering automatically any procedure of
measurement like that indicated {\em (\cite{Bohr35}, page 696).}
\eq

Une anecdote concernant cet article {\bf doit} être évoquée, car elle
circule en étant indûment attribuée à Schrödinger, les indications précises
ci-dessous rendent à Bohr ce qui lui appartient. Cet article donc n'est pas
d'une très grande clarté et lors de sa reproduction dans un livre consacré à
la mesure \cite{Whee83}, deux pages ont été inversées qui par malchance
gardent un semblant de sens, si bien que du temps s'est écoulé avant que
l'inversion soit découverte. Les pages de l'article original de
Physical Review 698, 699, 700, 701, sont devenues 147, 149, 148, 150 dans le
livre sur la mesure.

\subsection{La réponse forte de Schrödinger
aux sollicitations fortes de EPR.}

C'est avec la deuxième livraison de son article que Schrödinger commence à
répondre à EPR. Si on doutait de la force de l'argumentation EPR, on serait
surpris de celle de la réplique de  Schrödinger. Le chapitre 6, {\em La volte
face délibérée du point de vue épistémologique,} contredit en effet
directement les prémisses épistémologiques d'EPR (le pilier!) qui semblaient
pourtant aller de soi.

\bq
From this very hard dilemma [ce ne sont pas des variables déterminées mais
inconnues, les variables ne sont pas brouillées]  the reigning
doctrine rescues itself, or us, by having recourse to epistemology. We are
told that no distinction is to be made between the state of a natural object
and what I know about it \ldots Actually -so they say- there is intrinsically
only awareness, observation, measurement.

For we must now explicitely NOT relate our thinking any longer to any other
kind of reality or to a model {\em [page 157 ; les pages indiquées sont
celles de la référence \cite{Whee83}]}.
\eq

On appréciera l'humour de Schrödinger qui expose la doctrine mais la
manipule avec des pincettes! Mais la volte-face est en effet considérable,
on s'en apercevra mieux par les conséquences qui s'ensuivent mais qui déjà
se profilent et concernent le statut de la fonction d'onde, le rôle de la
mesure, de l'observation de l'observateur : tous ces problèmes dont en fait,
la MQ ne s'est pas encore dépêtrée aujourd'hui.

Cela apparaît mieux encore avec le chapitre 7. Avec son titre, ``{\em La
fonction $\Psi$ comme catalogue des valeurs possibles.}'' Schrödinger
définit ainsi le statut de la fonction d'onde\footnote{ Le choix du
caractère abstrait de la fonction d'onde n'est pourtant pas sans conduire à
des difficultés. Citons par exemple Bohr \cite{Bohr35} page 697 : ``{\em the
diffraction by the slit of the plane wave giving the symbolic representation
of its state will imply\ldots}''. Comment une onde symbolique peut-elle être
diffractée par une fente réelle? ou bien la représentation symbolique de la
fente (macroscopique) est-elle la fente elle même?}. Mais cela entraîne la
question suivante car un catalogue n'existe {\bf que pour} un observateur
qui le consultera un jour ou l'autre. Et surtout :

\bq[The abrupt change by measurement] is precisely {\em the} point that demands
the break with naive realism. For {\em this} reason one can {\em not} put
the $\psi$-function directly in place of the model or of the physical thing.
And indeed not because one might never dare impute abrupt unforeseen
changes to a physical thing or to a model, but because in the realism point
of view observation is a natural process like any other and cannot {\em per se}
bring about and interruption of the orderly flow of natural events {\em [page
158]}.
\eq

Ce qui fait apparaître une distinction entre la physique proprement dite et la
mesure, cette dernière débordant du cadre de la physique. C'est bien cette volte
face et elle seule (l'invocation de la complémentarité par Bohr est
inefficace) qui permet de désamorcer la question EPR avec le terme ii) de
l'alternative : il n'y a plus d'interaction après l'interaction car on n'est
plus dans la physique, ce qui  évolue, ce sont les connaissances :

\bq
If two separated bodies, each by itself known maximally, enter a situation
in which they influence each other, and separate again, then there occurs
regularly that which I have just called {\em entanglement} of our knowledge
of the two bodies.{\em [page 161]}. \eq

C'est là qu'est pour la première fois introduit le mot (Verschränkung,
entanglement, enchevêtrement/intrication). Ce mot est aujourd'hui largement
utilisé, on parle de ``systèmes intriqués'',
mais ce dont parle Schrödinger, ce n'est pas des systèmes mais bien de la
connaissance que l'on a d'eux. La différence n'est pas minime, c'est tout
simplement la volte-face épistémologique sans laquelle on retombe sous le
coup de EPR, on le verra à la fin de cet article.

Schrödinger envisage, comme il se doit, l'utilisation de dispositifs
d'enregistrement automatiques, ce qui met encore mieux en évidence le fond
du problème :

\bq Now how do things stand after automatically completed measurements? We
posess, afterwards same as before, a maximal expectation-catalog for the
total system. The recorded measurement result is of course not included
therein. As to the instrument the catalog is far from complete, telling us
nothing at all about where the recording pen left its trace (remember that
poisoned cat!). What this amounts to is that our knowledge has evaporated
into conditional statements : {\em if} the mark is at line 1, {\em then}
things are this and so for the measured object, {\em if} it is at line 2
{\em then} such and such, {\em if} at line 3, {\em then} a third\ldots
etc {\em [page 161].}
\eq

 On retrouve là aussi, intrication des connaissances et pas des systèmes. On
remarquera la référence au chat et c'est la seule dans tout l'article. Elle
est légitime et on notera que là non plus, il n'est pas question d'observation
d'états de superposition, mais bien de corrélations conditionnelles : si le
chat est vivant l'atome n'est pas désintégré, s'il est mort oui.

Mais il faut bien revenir à la physique et à l'objet :
\bq
From this amalgamation the object can again be separated out only by the
living subject actually taking cognizance of the result of the measurement.
Some time or other this must happen if that which has gone on is actually to
be called a measurement -however dear to our hearts it was to prepare the
process throughout as objectively as possible\ldots {\em not until this
inspection}, which determines the disjunction does anything discontinuous or
leaping take place. One is inclined to call this a {\em mental} action, for
this object is already out of touch, is no longer physically affected : what
befalls  it is already past {\em [page 162].}
\eq

L'essentiel est accompli : le recours nécessaire à la conscience.

Il est utile
cependant de décrire la fin de l'article de Schrödinger : les chapitres 11 à
15 parus dans la troisième livraison et d'abord :
 {\em 11) Résolution de l'enchevêtrement/intrication.
Résultat dépendant de l'intention de l'expérimentateur.}

Jusqu'à ce chapitre, l'intrication traitée par Schrödinger était celle des
informations concernant un système microscopique et un instrument de mesure
après que ceux-ci soient entrés en interaction et aient terminé cette
interaction. Cette fois, il repasse au cas général de deux systèmes sans
rien préciser de leur nature, il se rapproche ainsi des systèmes I et II de
EPR qui vont être explicitement évoqués dans le chapitre suivant {\em 12) Un
exemple}. Ici, dans {\em 11)}, il insiste sur les conséquences logiques de
la doctrine en examinant selon le schéma décrit ci-dessous sur les
possibilités d'établir le catalogue d'un système à partir de mesures sur
l'autre. Un plan de mesures sur B peut conduire à un catalogue maximal sur B
et qui le devient en même temps sur A grâce aux conditions ``si ceci sur B
alors cela sur A'' ; conditions qui sont entièrement résolues et donc
éliminées dès lors que les résultats des mesures (sur B) sont connues. Mais
des plans de mesure différents peuvent être conduits sur B, à la volonté de
l'expérimentateur. Ces plans différents sur B conduisent à des catalogues
sur A différents. Si l'un des plans est effectivement conduit et mène à un catalogue
maximal/complet, et si l'autre est aussi conduit et mène lui aussi à un
catalogue maximal, on s'attendrait à ce que ces catalogues sur A obtenus
sans toucher à A et tous les deux maximals soient donc identiques. Il n'en
est rien et il peut même arriver qu'ils n'aient rien en commun.

C'est déjà précisément l'argument EPR même si celui-ci ne sera présenté
explicitement par Schrödinger qu'au chapitre suivant : {\em 12) Un exemple} ;
n'y revenons pas, nous, puisque nous avons déjà examiné l'article EPR.

Avec le chapitre  {\em 13) Continuation de l'exemple : toutes les mesures sont
sans équivoque enchevêtrées}, Schrödinger rappelle d'abord que selon
beaucoup de ses amis, la question ``Quelle réponse {\em aurait} été donné {\em
si} l'expérimentateur avait \ldots'' n'a rien à voir avec une vraie mesure et
n'a donc pas lieu d'être posée selon le nouveau point de vue
épistémologique. Il semble, une fois de plus, que Schrödinger ne se
satisfasse pas complètement de ce point de vue. Il s'étonne de nouveau de
l'intrication généralisée et s'interroge sur le mécanisme inattendu qui
couple toutes les variables des catalogues quand les deux systèmes ont
interagi :

\bq
For {\em every} measurement on the ``small'' [system], the numerical result
can be procured by a suitably arranged measurement on the large one, and
each measurement on the large stipulates the result that a particular
measurement on the small would give or has given (\ldots only the virgin
measurement on each system counts) {\em [page 165]}.
\eq
On doit noter l'insistance pour Schrödinger  sur le ``catalogue des
valeurs'' plutôt que l'état du système (voir encore dans la conclusion
ci-dessous).

Avec le chapitre {\em 14) La dépendance temporelle de l'enchevêtrement.
Considération du rôle spécial du temps}, Schrödinger note que les valeurs
des variables varient en général avec le temps mais que tout le formalisme
de la MQ concerne la valeur de ces variables à un instant parfaitement
défini, quels que soient la durée des mesures et le temps que peut attendre
l'observateur pour en prendre connaissance. Schrödinger s'interroge pourtant
sur le fait que la valeur numérique du temps suppose une observation et donc
d'une certaine façon l'intervention d'une horloge.

L'article se termine avec le chapitre {\em 15) Loi naturelle
ou moyen de calcul} :

Après avoir évoqué la difficulté de rendre la MQ compatible avec la
relativité, de traiter le champ électromagnétique\ldots Schrödinger revient
sur l'essentiel de son intervention, et qui est aussi l'essentiel de EPR,
dans un résumé remarquable :

\bq
The remarkable theory of measurement, the apparent jumping around of the
$\psi$-function, and finally the ``antinomies of entanglement'', all derive
from the simple manner in which the calculation methods of quantum mechanics
allow two separated systems conceptually to be combined together into a
single one; for which the methods seem plainly predestined. When two systems
interact, their $\psi$-functions, as we have seen, do not come into
interaction but rather they immediately cease to exist and a single one, for
the combined system, takes their place. It consists, to mention this
briefly, at first simply of the {\em product} of the two individual
functions ; which, since the one function depends on quite different
variables from the other, is a function of all these variables, or ``acts in
a space of much higher dimension number'' than the individual functions. As
soon as the systems begin to influence each other, the combined function
ceases to be a product and moreover does not again divide up, after they
have again become separated, into factors that can be assigned individually
to the systems. Thus one disposes provisionally (until the entanglement is
resolved by an actual observation) of only a {\em common} description of the
two in that space of higher dimension. This is the reason that knowledge of
the individual systems can decline to the scantiest, even to zero, while
that of the combined system remains continually maximal. Best possible
knowledge of the whole does {\em not} include best possible knowledge of its
parts -and that is what keeps coming back to haunt us {\em [page 167]}.
\eq

 et Schrödinger conclut finalement :
\bq The simple procedure provided for this by the non relativistic theory is
perhaps after all only a convenient calculational trick, but one that
today, as we have seen, has attained influence of unprecedented scope over
our basic attitude toward nature {\em [page 167]}.
\eq

Schrödinger est ici très loin de la satisfaction dogmatique de l'école de
Copenhague, mais, contraint et forcé il semble accepter ici, au contraire
d'Einstein, la volte-face épistémologique de l'abandon du réalisme. Il ne
s'en satisfera jamais.

\section{Pendant soixante-quinze ans : observation et conscience. }

Clairement revendiqué, honteusement accepté ou, sans contrepartie,
dédaigneusement ignoré, le recours à la conscience comme clef de voûte de la
MQ par son rôle dans la mesure est une constante qu'on va retrouver jusqu'à
nos jours. En voici quelques exemples, le premier antérieur à EPR : ce n'est
pas Schrödinger qui a inventé ce recours!

{\bf 1)} 
Au congrès Solvay de 1927, dans
la session ``Electrons and photons'', l'exposé ``Quantum mechanics''
de Born et Heisenberg présente le point de vue de Göttingen-Copenhague
et commence en insistant :
\bq
The new mechanics is based on the idea that atomic physics is essentially
different from classical physics on account  of the existence of
discontinuities {\em \cite{Mehr75}(page 146).}
\eq
On pourrait penser en termes de discontinuités physiques, mais
dans la discussion qui suit (page 149) Heisenberg s'oppose à Dirac et
explicite le rôle de ``l'observation'':
\bq
Quantum theory, he [Dirac] said, describes a state by a time-dependent wave
function $\psi$ which can be expanded at a given time {\em t} in a series
containing wave functions $\psi_n$ with coefficients $c_n$. The wave
functions $\psi_n$ are such that they do not interfere at an instant {\em t}
\textgreater {\em t$_1$}. Now Nature makes a choice sometimes later and decides in
favour of the state $\psi_n$ with the probability $|c_n|^2$. This choice
cannot be renounced and determines the future evolution of the state.
Heisenberg opposed this point of view by asserting there was no sense in
talking about Nature making a choice, and that it was our observation that
gives us the reduction to the eigenfunction. What Dirac called a ''choice of
Nature'', Heisenberg preferred to call ``observation'', showing his
predilection for the language he and Bohr had developed together.
\eq

{\bf 2)} On continue, après 1935 cette fois, en 1939, par un livre remarquable à bien
des égards : il est un des livres sur la MQ écrit en français et traduit
et reconnu par la communauté anglophone. La préface comme le livre sont
d'une grande clarté et s'y trouvent développés les éléments formels utiles
au traitement de la mesure, cas pur et mélange statistique, opérateurs et
matrices statistiques, trace partielle. Il traite comme EPR puis Schrödinger
du cas de deux systèmes I et II etc\ldots Cet ouvrage est préfacé par Paul
Langevin, connu pour ses positions plutôt du côté du
réalisme et du matérialisme, admirateur de A. Einstein, mais pour qui le
recours à la conscience semble être devenu un fait d'expérience.

\bq L'acte d'observation se trouve analysé ici de manière particulièrement
aiguë et les caractères essentiels de la nouvelle physique y ressortent avec
une entière clarté à travers les deux étapes du changement de la fonction
d'onde par couplage du système observé avec l'appareil de mesure et par
l'intervention de l'observateur qui prend conscience du résultat et fixe
ainsi la nouvelle fonction d'onde consécutive à la mesure en intégrant
celle-ci à l'information antérieurement acquise
{\em  \cite{Lond39} Préface de Paul Langevin, page 5.} \eq

Les auteurs, F. London et E. Bauer, détaillent évidemment plus et
distinguent :
\bq a) Les transformations réversibles que l'on peut appeler aussi ``causales''.
{\em [dirait-on aujourd'hui déterministes? qui correspondent à l'évolution
de la fonction d'onde selon l'équation de Schrödinger]}. Elles transforment
un cas pur en un cas pur {\em pages 38-39}.
\eq

et

\bq b) Les transformations irréversibles que l'on pourrait appeler aussi
``acausales''. Celles-ci ne se réalisent qu'à l'occasion d'un contact du
système en question (I) avec un autre système (II). Le système total
embrassant les deux systèmes (I+II) subit encore une transformation
réversible tant que l'ensemble I+II est isolé.
 Mais si nous fixons notre attention [{\em trace partielle}] sur le
système I, ce dernier subira une transformation irréversible : s'il était
dans un état pur avant le contact, il sera en général transformé en un
mélange {\em [page 39]}.

Celle-ci [{\em la mesure}] est achevée seulement lorsqu'on a {\em observé}
la position de l'aiguille\ldots [{\em ce}] qui donne à l'observateur le
droit de choisir entre les différentes composantes du mélange prévus par la
théorie, de rejeter celles qui ne sont pas observées et d'attribuer
dorénavent à l'objet une nouvelle fonction d'onde, celle du cas pur qu'il a
trouvé {\em [page 41] revoir la présentation de la MQ dans le présent
article aux pages 5 à 7}.
\eq

On notera comment la transformation irréversible est obtenue en ``fixant
notre attention''!

{\bf 3)} On retrouve presque la même expression chez L. de Broglie avec ce texte
écrit en 1950-51 mais publié 30 ans plus tard :

\bq
Les transformations irréversibles correspondant à des processus non soumis
au déterminisme se produisent au moment des interactions de mesure
précédemment analysées. L'interaction du système étudié 1 avec l'appareil de
mesure 2 correspond à une évolution déterminée et réversible de l'état
global du système 1+2 jusqu'à ce que la constatation macroscopique de l'état
individuel du système 2 par l'observateur vienne interrompre cette évolution
en attribuant au système 1 un nouvel état par un processus qui n'est ni
réversible ni [même] causal [dans l'interprétation actuelle] {\em
\cite{Brog82}.}
\eq

{\bf 4)} Avec Wigner (en 1961 mais réédité en 62, 67 et 83), on est dans
l'affirmation triomphante :

\bq When the province of physical theory was extended to encompass
microscopic phenomena, through the creation of quantum mechanics, the concept
of consciousness came to the fore again : it was not possible to formulate
the laws of quantum mechanics in a fully consistent way without reference to
the consciousness. All that quantum mechanics purport to provide are
probability connections between subsequent impressions (also called
``apperceptions'') of the
consciousness, and even though the dividing line between the observer,
whose consciousness is being affected, and the observed physical object
can be shifted towards the one or the other to a considerable degree, it
cannot be eliminated {\em \cite{Wign61} et \cite{Whee83} page 169.}

{\bf It follows that the being with a consciousness must have a different
role in quantum mechanics than the inanimate measuring device...}
 {\em \cite{Wign61} et \cite{Whee83} page 177.}
\eq

{\bf 5)} Pour Wheeler, reprenant à son compte les arguments de Bohr, on
semble d'abord dans la réalité objective (l'amplification irréversible),
mais on retrouve un rôle final pour l'observateur avec le recours à
l'observation :
\bq
In today's words Bohr's point -and the central point of quantum theory- can
be put into a single simple sentence. `No elementary  phenomenon is a
phenomenon until it is a registred (observed) phenomenon''. {\em [et plus
loin :]}

 For a proper way of speaking we recall once more that it makes no sense to
talk of the phenomenon until it has been brought to a close by an
irreversible act of amplification : ``No elementary  phenomenon is a
phenomenon until it is a registred (observed) phenomenon''. {\em \cite{Whee81}
et \cite{Whee83} pages 184 et 192}.\eq

Et puis Wheeler, dans ces mêmes textes, au début des années 80, rapproche
finalement et humoristiquement la MQ de l'art et même de la foi :

\bq How can one contemplate indeterminism, complementarity and ``phenomenon''
without being reminded of the words of Gertrude Stein about modern art? ``
It looks strange and it looks strange and it looks very strange ; and then
suddenly it does not look strange at all and you can't understand what made
it look strange in the first place {\em \cite{Whee81} et
\cite{Whee83} page 185}.\eq

{\bf 6)} On arrive maintenant avec Zurek dans la période contemporaine. Dans 
l'introduction de l'article qui justement présente la décohérence, Zurek est
assez ambigu et évasif   (on reprend et on
complète la citation du début d'article) :

\bq
Given almost any initial condition the universe described by
$\ket{\Psi}$ evolves
into a state that simultaneously contains many alternatives never seen to
coexist in our world.
Moreover, while the ultimate evidence for the choice of one such option
resides in our elusive ``consciousness'', there is every indication that the
choice occurs long before consciousness ever gets involved {\em \cite{Zure91} page
36.}\eq

D'un côté Zurek semble donc devoir reconnaître le rôle ultime de la
conscience, mais ignorant toutes les questions que cela a soulevées depuis si
longtemps, il tente de l'autre de se débarrasser de cet encombrant recours
par le simple appel à du bon sens.

{\bf 7)} Dans son traité de MQ (1993) Asher Peres reprend l'expérience (devenue paradoxe) du
chat de Schrödinger en même temps qu'il (re)pose la question du statut de la
fonction d'onde -il tranche en fait contre son statut de réalité-. Il
retrouve alors curieusement l'expression de Schrödinger mais élude par
contre sa conclusion concernant la nécessité de la
volte-face épistémologique et donc du rôle de l'observateur (l'action mentale).

\bq Clearly, the ``cat paradox'' arises because of the naive assumption that
the time evolution of the state vector $\psi$ represents a physical process
which is actually happening in the real world.
In fact, there is no evidence whatsoever that every physical system has at
every instant a well defined $\psi$ (or a density matrix $\rho$) and that the
time dependence of $\psi(t)$, or of $\rho(t)$, represents the actual evolution
of a physical process. In a strict interpretation of quantum theory, these
mathematical symbols represent {\em statistical information} enabling us to
compute the {\em probabilities} of occurrence of specific
events {\em \cite{Pere93} page 373}.\eq

et un peu plus loin :

\bq There have been many attempts to save the objectivity of the wave function
by arguments such as : ``Nobody has ever {\em seen} a cat in state (12,1)
[état de superposition], but this is only because the mere {\em observation}
of the cat causes $\psi$ to jump into either [vivant ou mort] states'' (this
jump is called a {\em collapse} of the wave function). There is nothing
formally inconsistent in this scenario, but it is nevertheless incredible,
as it implies a powerful and mysterious interaction between the brain of the
observer and the entire body of the cat. {\bf A measurement, after all,
is not a supernatural event. It is a physical process, involving ordinary
matter, and whatever  happens ought to be analysed by means of the ordinary
physical laws.} {\em[page 374]}.\eq

{\bf 8)} Et puis aujourd'hui (oct-nov 2002) avec B. d'Espagnat \cite{Espa02} 
\bq
Les sciences ont beau ne porter que sur l'expérience humaine, jusqu'à
l'avènement de la MQ leurs principes constitutifs pouvaient s'énoncer dans un
langage objectiviste, c'est-à-dire comme s'ils portaient sur l'être-en-soi.
Aujourd'hui, c'est la physique elle-même qui -via la MQ- nous fait toucher
du doigt la difficulté qu'il y a à ériger ses découvertes en descriptions
fidèles d'une réalité de base indépendante de nous-mêmes\ldots 
{\bf Quant aux concepts quantiques, leur nature est d'être attachés non pas à des
entités, mais bien à des opérations -humaines- de préparation et
d'observation.} {\em [\cite{Espa02} page 75. ]}
\eq

\section{Les développements récents.}

Des éléments nouveaux sont à considérer aujourd'hui. D'abord, la
démonstration expérimentale, au moins la présomption forte, que les
paramètres cachés {\bf dans le passé commun} (le premier terme de
l'alternative EPR) ne peuvent expliquer le fonctionnement de la ``mesure''
quantique.
On verra par exemple les expériences d'A. Aspect et al \cite{Aspe82}. On ne
discutera pas de ces questions ici : d'une part, ce n'est pas du tout
l'objet de cet article, d'autre part il est peut-être possible de se
convaincre et c'était probablement le cas d'Einstein en 1949 (voir plus
loin) que là ne peut pas être la solution.

Les deux éléments les plus directement liés au sujet sont d'une part les
développpements expérimentaux et théoriques qui s'articulent autour de
l'optique quantique, les manipulations d'atomes et de champs cohérents dans
une cavité. D'autre part, un concept a été introduit, la décohérence, dont
les objectifs sont de traduire plus formellement, on verra plus loin comment,
les intuitions de l'école de Copenhague
({\em ``This analysis is fully consistent with the Copenhagen description of  a
measurement''} écrit par exemple J.M. Raimond \cite{Raim01} page 566).

Nous présenterons d'abord la spécificité des ``états cohérents'' : la
possibilité de passer avec eux continûment du microscopique au
macroscopique. On présentera ensuite une expérience type associant
manipulations d'atomes et états cohérents. Cette expérience
schématique permettra de présenter les processus à l'\oe uvre au cours d'une
``mesure'' et d'expliciter le concept de décohérence. 
On s'appuiera pour cela sur le travail accompli dans ce domaine au
Laboratoire Kastler-Brossel de l'ENS et dont on a déjà beaucoup cité les
articles. Ces articles doivent être lus indépendamment de leur intérêt pour
les questions discutées ici.  Bien entendu, on ne rapportera, et
schématiquement, que ce qui concerne notre propos, le problème de la mesure
quantique et les corrélations EPR.

\subsection{États cohérents ou semi-classiques : une possibilité d'états
mésoscopiques.}

Si on se propose d'explorer le passage du microscopique au macroscopique, il
est intéressant de disposer d'un objet qui puisse lui-même exister dans les
deux champs et effectuer la transition de l'un à l'autre. Les états
cohérents ou semi-classiques sont de bons candidats pour cette fonction.

En électrodynamique classique, le champ existant dans une cavité peut être
décrit comme résultant de l'addition d'oscillateurs harmoniques dont les
fréquences sont les fréquences ``naturelles'' de la cavité. Avec
l'électrodynamique quantique, ce sont ces oscillateurs, les modes du champ,
qui sont quantifiés.

Pour chaque mode de fréquence angulaire $\omega$, on définit les opérateurs
de création $a^+$ et d'annihilation $a$. L'opérateur nombre (de photons de
ce mode) s'écrit alors $N = a^+a$.

Deux types d'état très différents sont alors à considérer 

- Les états $|n>$ qui sont les états propres de l'opérateur nombre :

\[ N |n> = a^+ a |n> = n|n> \]

- Les états cohérents \cite{Glau65} ou semi-classiques ou d'incertitude minimale,
qui sont les états propres de l'opérateur d'annihilation :

\[ a | \alpha > = \alpha | \alpha > \]

Les états $| \alpha > $  sont une combinaison linéaire des états $|n>$ :

\[ | \alpha > = exp (-1/2 | \alpha | ^2) \sum_{n=0}^{\infty}
\frac{\alpha^n}{(n!)^{1/2}} | n > \]

En conséquence, ils ne correspondent pas à un nombre donné de photons. Une
mesure de $n$ donnera différents résultats avec les probabilités :

\[ P (n) = \frac{| \alpha | ^{2n} exp (-|\alpha |^2)}{n!} \]

Les états $|n>$ sont stationnaires ; même si $n$ est grand, l'état
stationnaire $|n>$ ne peut approcher une onde classique.

les états $|\alpha>$ ne sont pas stationnaires et leur amplitude complexe
varie avec le temps :

\[ \ket{\alpha (t)} = \ket{\alpha e^{-i \omega t} } \]

C'est ce qui permet à un état $\ket{\alpha}$ de s'approcher d'une onde
classique autant que la mécanique quantique le permet, et elle le permet
d'autant mieux que $|\alpha|$ est grand. Ceci est illustré sur le schéma
ci-dessous où l'onde est représentée par son vecteur de Fresnel qui donne
son amplitude et sa phase. Le cercle hachuré de rayon {\em r} représente
l'incertitude sur l'une et l'autre. Le rayon {\em r} est d'autant plus petit que
$|\alpha|$ est grand.

On voit bien alors la possibilité d'atteindre un état mésoscopique voisin de
l'un ou de l'autre des champs microscopique ou macroscopique selon que
$|\alpha|$ est petit ou grand.

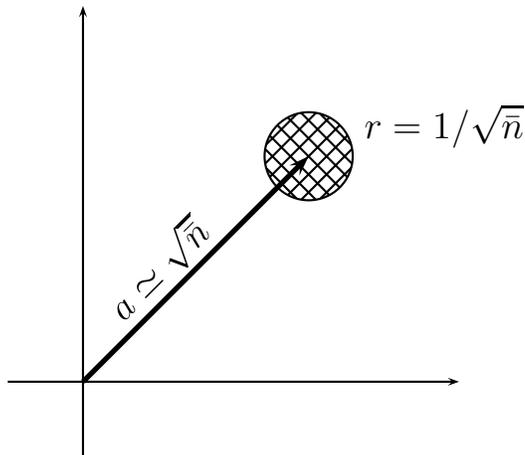
\vspace*{.5cm}

\hspace*{1.5cm}
\begin{pspicture}(5,5)
\psline{->}(-1,0)(5,0)
\psline{->}(0,-1)(0,5)
\psline[linewidth=.07]{->}(0,0)(3,3)
\pscircle[fillstyle=crosshatch](3,3){.6}
\rput{45}(1,1.5){{\Large $a\simeq\sqrt{\bar{n}}$}}
\rput{0}(4.8,3.4){{\Large $ r = 1/\sqrt{\bar{n}}$}}
\end{pspicture}

\vspace{7mm}

Dans la situation EPR, l'un des systèmes I ou II pourra voir ainsi son
caractère macro- ou micro- varier.

\subsection{Expérience : les éléments de base.}

L'expérience type (voir schéma ci-dessous) fait interagir un atome de
Rydberg circulaire (le système I) et un état cohérent présent dans une
cavité (le système II). L'atome doté d'une grande vitesse traverse la cavité
et c'est le mouvement de cet atome qui ``allume'' puis ``éteint''
l'interaction en introduisant puis retirant l'atome de la cavité et donc de
l'interaction avec le champ.

L'atome est ensuite lui-même soumis à une mesure par corrélation avec un
détecteur d'état par ionisation (système III).

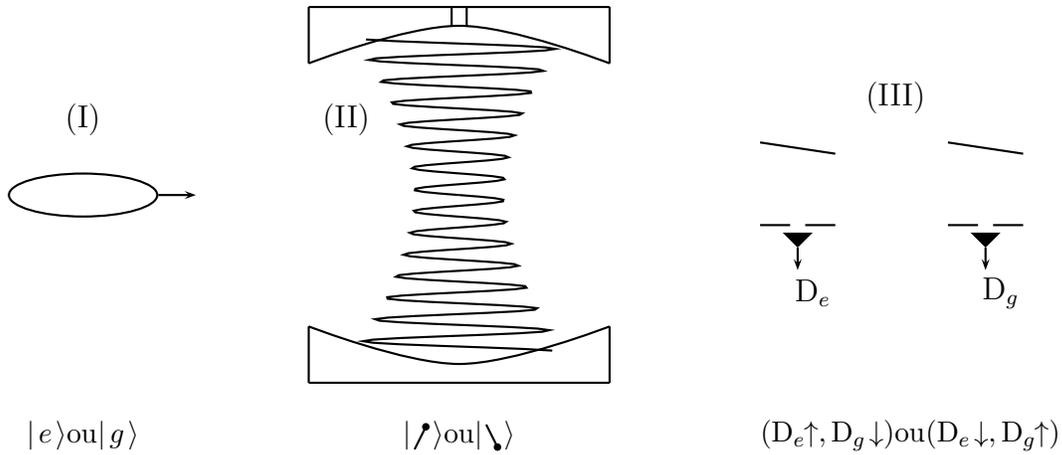
\begin{pspicture}(13,8.5)
\rput(1,3.5){

\psellipse(0,0)(1,.3)
\psline{->}(1,0)(1.5,0)
\rput(0,1){\large (I)}

\rput{0} (5,-2.5){
\psline(-2,0)(2,0)
\psline(-2,0)(-2,0.75)
\psline(2,0)(2,0.75)
\pscurve(-2,0.75)(0,.25)(2,0.75)
}

\rput{180} (5,2.5){
\psline(-2,0)(2,0)
\psline(-2,0)(-2,0.75)
\psline(2,0)(2,0.75)
\pscurve(-2,0.75)(0,.25)(2,0.75)
\psline(-0.1,0)(-0.1,.26)
\psline(0.1,0)(0.1,.26)
}

\rput{90}(5,0){
\psset{xunit=.04}
\psplot[plotpoints=100]{0}{51.75}{x 50 mul sin x 2 mul cos -1 mul 2 add mul .6 mul}
}

\rput{-90}(5,0){
\psset{xunit=.04}
\psplot[plotpoints=100]{0}{51.75}{x 50 mul sin x 2 mul cos -1 mul 2 add mul .6 mul}
}

\rput(3.5,1.){\large (II)}

\rput{0}(9,0){
\psline(0,.7)(1,.55)
\psline(0,-.4)(.4,-.4)
\psline(.6,-.4)(1.,-.4)
\pspolygon*(.3,-.5)(.7,-.5)(.5,-.7)
\psline{->}(.5,-.7)(.5,-1)
}
\rput{0}(11.5,0){
\psline(0,.7)(1,.55)
\psline(0,-.4)(.4,-.4)
\psline(.6,-.4)(1.,-.4)
\pspolygon*(.3,-.5)(.7,-.5)(.5,-.7)
\psline{->}(.5,-.7)(.5,-1)
}

\rput(9.7,-1.3){\large D$_e$}
\rput(12.2,-1.3){\large D$_g$}
\rput(10.8,1.3){\large (III)}

$\rput(0,-3.2){\ket{e} \mathrm{ou} \ket{g}}
\rput(5,-3.2){\ket{\chu} \mathrm{ou} \ket{\chd}}
\rput(11,-3.2){(\deu,\dgd) \mathrm{ou} (\ded,\dgu)}
$

}
\end{pspicture}

\vspace{3mm}

\begin{minipage}[t]{16cm}
\begin{multicols}{2}

L'atome de Rydberg circulaire peut être préparé dans trois états {\em i},
{\em g}, {\em e}, correspondant respectivement aux trois valeurs du nombre
quantique principal n= 49, 50, 51. On ne s'interessera ici qu'aux états e et
g. Cet atome, de rayon 1000 fois plus grand que dans son état fondamental et
d'autant plus grand que {\em n} est grand, a une interaction dipolaire avec
le champ cohérent de la cavité dont la fréquence est très voisine de la
transition e $\iff$ g à 52,1 GHz.

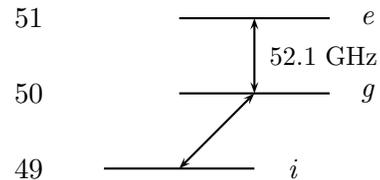
\begin{pspicture}(6,7)
\rput(1,2.5){

\rput(0,0){49}
\psline{-}(1,0)(3,0)
\rput(3.5,0){\em i}

\rput(0,1){50}
\psline{-}(2,1)(4,1)
\rput(4.5,1){\em g}

\rput(0,2){51}
\psline{-}(2,2)(4,2)
\rput(4.5,2){\em e}

\psline{<->}(2,0)(3,1)
\psline{<->}(3,1)(3,2)
\rput[l](3.2,1.5){\small 52.1 GHz}
}

\end{pspicture}
\end{multicols}
\end{minipage}
\begin{multicols}{2}

Par effet d'indice de réfraction, celui qui serait obtenu par l'introduction
d'un matériau transparent dans la cavité, le passage de l'atome dans la
cavité conduit à une rotation du vecteur de Fresnel du champ cohérent, dans
un sens si l'atome est dans l'état e, dans le sens opposé s'il est dans
l'état g.

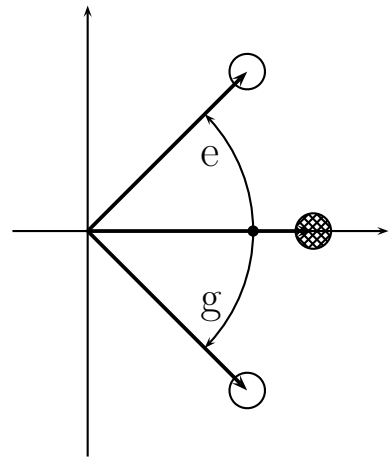
\begin{pspicture}(3,4)
\rput[l](2,2.5){
\psline{->}(-1,0)(4,0)
\psline{->}(0,-3)(0,3)
\psarc{->}(0,0){2.2}{0}{45}
\psarc{<-}(0,0){2.2}{-45}{0}
\pscircle*(2.2,0){.075}
\rput[l]{0}(1.5,1.) {\LARGE {e}}
\rput[l]{0}(1.5,-1.){\LARGE {g}}

\rput[3,3]{-45}(0,0){
\psline[linewidth=.05]{->}(0,0)(3,0)
\pscircle(3,0){.25}
}

\rput[3,3]{0}(0,0){
\psline[linewidth=.05]{->}(0,0)(3,0)
\pscircle[fillstyle=crosshatch,hatchsep=.05](3,0){.25}
}

\rput[3,3]{45}(0,0){
\psline[linewidth=.05]{->}(0,0)(3,0)
\pscircle(3,0){.25}
}
}
\end{pspicture}

\end{multicols}

L'atome est ensuite ionisé, au passage dans le premier détecteur (marqué
D$_e$) s'il est dans l'état e, au passage dans le second (marqué D$_g$,
où le champ est plus fort) s'il est dans l'état g.

Dans une première expérience schématique, les atomes (le système I) sont
préparés dans l'état e ou dans l'état g. Dans la situation ainsi décrite, les
états des trois systèmes (I,II,III) se trouvent corrélés sans surprise ni
résultat aléatoire :

{\em ou bien} on envoie $\ket{e}$ et il s'ensuit que le champ est porté
dans l'état $\ket{\chu}$ et le détecteur D$_e$ déclenche (on notera $\deu$), 

{\em ou bien} on
envoie $\ket{g}$, le champ est porté dans l'état $\ket{\chd}$, le
détecteur D$_g$ déclenche (on notera $\dgu$).

Pas de doute non plus que c'est la nature
initiale de l'atome $\ket{e}$ ou $\ket{g}$ qui a déterminé, d'abord le
basculement du champ, puis le déclenchement du détecteur approprié au cours
du trajet de cet atome.

On peut représenter ces deux expériences par les deux schémas :

\begin {equation}
\ket{e} \Longrightarrow \ket{e}\ket{\chu} \Longrightarrow
\ket{\chu}(\deu,\dgd)
\end{equation}
\begin {equation}
\ket{g} \Longrightarrow \ket{g}\ket{\chd} \Longrightarrow
\ket{\chd}(\ded,\dgu) 
\end {equation}

On a fait disparaître à droite la représentation de l'atome puisqu'{\em après}
déclenchement de l'un ou l'autre des détecteurs l'atome est ionisé et on ne
s'intéresse plus à son état.

On doit garder à l'esprit, à la lecture de ces schémas et de ceux qui
suivent, la succession des événements dans le temps et dans l'espace.
L'atome se déplace à vitesse finie, le champ est confiné dans la cavité et
les détecteurs situés à quelques centimètres de celle-ci.

\subsection{Expérience : on introduit un état de superposition.}
Pour faire intervenir maintenant des états quantiques corrélés, il faut
introduire un état de superposition sur le système (I). Pour cela, les
atomes d'abord préparés dans l'état e sont ensuite portés dans un état de
superposition e+g, par {\bf un dispositif non représenté, avant de pénétrer dans
la cavité}. Après le passage dans la cavité, on est donc passé d'un état de
superposition à un état de superposition d'états corrélés, des paires EPR
pourrait-on dire. Mais il n'y a pas encore eu mesure proprement dite, il n'y
a pas eu irréversibilité (ni intervention de conscience!) car en principe,
on peut aussi bien décider de mesurer l'atome pour avoir l'état du champ
qu'à l'inverse, mesurer le champ pour avoir l'état de l'atome. C'est le
premier cas qu'on va (d'abord) considérer : la mesure sera accomplie sur
l'atome par les détecteurs D$_e$ ou D$_g$.

Avant de pénétrer dans les détecteurs, on a maintenant le schéma, unique
cette fois (une seule préparation, un seul résultat jusque-là) :

\begin{equation} 
\ket{e} \Longrightarrow \ket{e}+\ket{g} \Longrightarrow
\ket{e}\ket{\chu}+\ket{g}\ket{\chd}
\end{equation} 

Le champ dans la cavité est mésoscopique, mais dans les expressions
ci-dessus, il a été traité comme microscopique : les superpositions sur les
états de l'atome sont corrélées aux superpositions du champ (intrication,
entanglement, Verschränkung). Avec l'étape suivante, on interrompt cette
chaîne de corrélations pour passer au processus de mesure proprement dite
associée à l'intervention des instruments macroscopiques que sont les
détecteurs à ionisation D$_e$ et D$_g$. On applique les postulats et
l'aspect probabiliste apparaît en même temps que le cas pur devient un
mélange :
\begin{equation}
\ket{e}\ket{\chu}+\ket{g}\ket{\chd}\Longrightarrow
\begin{cases}
\ket{\chu}(\deu,\dgd) & (e)  \\
\mathrm{ou} & \\
\ket{\chd}(\ded,\dgu) & (g) \\
\end{cases}
\end{equation}

On a chacune de ces possibilités (e ou g) avec la même probabilité et
avec la même préparation d'atome initiale $\ket{e}$.

Supposons maintenant que nous n'ayons pas interrompu la chaîne des
corrélations à l'entrée du détecteur ou encore que celui-ci ait été traité
comme microscopique. On aurait alors à la place du passage de (3) à (4), le
passage à

\begin{equation}
\ket{e} \Longrightarrow \ket{e}+\ket{g} \Longrightarrow
\ket{e}\ket{\chu}+\ket{g}\ket{\chd} \Longrightarrow 
\ket{\chu}
(\deu,\dgd)+
\ket{\chd}(\ded,\dgu)
\end{equation}

Si le passage de l'état de superposition des détecteurs dans (5) au tirage
de l'un ou l'autre de ces états dans (4) n'est rendu légitime que par le
caractère macroscopique du détecteur, il faut insister sur la part
d'arbitraire de cette distinction entre micro- et macroscopique (nous
restons au c\oe ur du problème de la mesure). C'est pour tenter de réduire
cette part d'arbitraire que le concept
de décohérence a été introduit.

\subsection{Le principe de la décohérence.
}

C'est le passage de l'expression (3) à l'expression (4) d'une part ou à
l'expression (5) d'autre part que la décohérence va détailler et tenter de
régler.

Deux modifications mineures formelles sont introduites :
\begin{quote}
i) Pour  simplifier l'expression, on va supposer que l'ensemble des détecteurs
D$_e$  et D$_g$ 
est réduit à un seul détecteur
D$_g$ qui, soit déclenche indiquant que l'atome ``était'' dans l'état g et
que le champ ``est'' désormais dans l'état bas, soit qu'il n'a pas déclenché,
que l'atome ``était'' dans l'état e et que le champ ``est'' désormais dans
l'état haut.

ii) Le mélange e et g est effectué avec les poids explicites $\alpha$ et $\beta$
(au lieu de 1 et 1) afin de mieux garder la trace de l'origine des
différents termes dans les développements qui suivent.
\end{quote}

Immédiatement après (3), mais en introduisant la représentation du détecteur
{\em avant} son interaction,
l'état inital est un état (partiellement) factorisé :

$$\ket{\phi_i}=( \alpha \ket{e,\chu} + \beta \ket{g,\chd})\ket{\dgd}$$

Après l'interaction, il n'est plus factorisé (il s'agit d'une simple
réécriture de (5)) :

$$\ket{\phi_c}=\alpha \ket{\chu}\ket{\dgd} + \beta \ket{\chd}\ket{\dgu} $$

On passe alors à la représentation en matrice densité $ \rho_c$ de
$\ket{\phi_c}$ qui a la vertu de
permettre de décrire aussi bien les cas purs que les mélanges statistiques
et qui est donc l'instrument adapté au passage de (5) à (4), c'est-à-dire
à la description de la mesure.

$$ \rho_c=\ket{\phi_c}\bra{\phi_c}=$$

$$
 |\alpha|^2 \ket{\chu}\bra{\chu} \ket{\dgd}\bra{\dgd} +
  |\beta|^2 \ket{\chd}\bra{\chd} \ket{\dgu}\bra{\dgu}
 $$

$$
+
\alpha\beta^* \ket{\chu}\bra{\chd} \ket{\dgd}\bra{\dgu}
 + \alpha^*\beta \ket{\chd}\bra{\chu} \ket{\dgu}\bra{\dgd}$$

On n'a pas encore introduit de mesure mais seulement une corrélation
supplémentaire des systèmes I et II déjà corrélés avec le système III des
détecteurs : il s'agit d'une réécriture dans le formalisme de la matrice
densité de l'expression (5) qui représente un cas pur.

Dans les quatre éléments de la matrice densité, les éléments diagonaux 
(première ligne),
dont les poids sont $|\alpha|^2$ et $|\beta|^2$,  sont simples à
interpréter car le détecteur et le champ sont dans des états correspondants.
Les éléments non diagonaux, dits d'interférence, ne le sont pas puisque les
états du champ et ceux du détecteur sont croisés. Ce sont justement ces
termes qui, par le couplage avec l'environnement (un système IV,
macroscopique lui), vont être conduits très rapidement à zéro : c'est le
principe de la décohérence, son intervention spécifique.

Avec ses termes d'interférence, la matrice densité était celle d'un cas pur,
maintenant, sans eux, elle est celle d'un mélange qui correspond dans la
pratique {\bf ou bien} à l'observation du détecteur dans l'état $\dgd$ {\bf
ou bien} dans l'état $\dgu$, c'est à dire au passage à l'expression (4). En
terme d'observations, la répétition de l'expérience voit ces résultats
($\dgd$ ou $\dgu$) trouvés avec les probabilités correspondantes
$|\alpha|^2$ et $|\beta|^2$.

Ce que sont les termes diagonaux et ceux d'interférence dépend du choix de
la base (une base construite avec des combinaisons linéaires orthogonales de
$\dgd$ et $\dgu$ donnerait un tout autre résultat), mais ce choix n'est
qu'apparemment arbitraire : la base qui doit être choisie est celle où les
détecteurs concrets (et le champ) sont dans les états effectivement
observés, ici $\dgd$ et $\dgu$. On peut dire que c'est là que se manifeste
un rapport étroit avec la ``complémentarité'' de Bohr : les expériences
différentes pour mesurer des variables différentes et par exemple
complémentaires vont conduire à des descriptions théoriques différentes et
finalement à des résultats différents. Cela ne constitue cependant pas un
fondement de cette complémentarité.

Peut-être n'est-il pas complètement légitime de parler d'un {\em processus}
de décohérence, cela place d'emblée celle-ci dans le domaine de la physique
et pas dans celui de la représentation abstraite, cependant la décohérence a
une durée. On verra plus bas que cette durée est d'autant plus courte que
les états diagonaux considérés sont plus éloignés et donc aussi qu'ils sont plus
macroscopiques.

On a supposé là que le formalisme de la décohérence \ldots  était cohérent et
efficace sans du tout en examiner le fonctionnement. C'est au mieux ce qu'il
peut être et on verra plus bas quelles en sont les conséquences. On risquera
un seul commentaire sur ce concept : on ne peut passer d'un cas pur, une
fonction d'onde, à un mélange, que si l'environnement est lui-même déjà un
mélange. La décohérence si elle fonctionne ne peut donc expliquer en quelque
sorte que la contagion du mélange mais pas la naissance initiale de ce
mélange. On ne s'interesse ici à la décohérence que par rapport à
l'alternative EPR ; on supposera qu'elle fonctionne pleinement comme elle le
prétend.

\subsection{La décohérence en compétition.}
À la sortie de la cavité, on l'a dit plus haut, le système corrélé atome-champ
$$ \ket{\phi_i}= \ket{e,\chu}+\ket{g,\chd}$$
est susceptible de subir aussi bien une mesure sur l'état e ou g de l'atome
(qui détermine l'état correspondant du champ) qu'une mesure de l'état $\chu$
ou $\chd$ du champ (qui détermine celui de l'atome.) Dans le premier cas, on
suppose que le champ n'est soumis à aucune
interaction avec l'environnement, dans le second, c'est l'atome qui doit
rester épargné. C'est le premier cas qu'on a traité jusqu'à maintenant. La
décohérence était à l'\oe uvre seulement pour les détecteurs D.

Le champ cohérent est en fait soumis lui aussi à l'environnement par la
``fuite'' des photons d'une part et par l'absorption des photons par les
défauts des miroirs. Si bien qu'on peut dire que les interactions des deux
sous-systèmes (l'atome et le champ) avec leur environnement, les deux
décohérences, sont en compétition et ce sont leurs temps caractéristiques
qu'il faut comparer en tenant compte du décalage temporel dû au transit de
l'atome depuis la sortie de la cavité jusqu'aux détecteurs D (quelques
centimètres parcourus à 400m/s).

Examinons les différents temps qui interviennent :
\begin{quote}
T$_r$ le temps de vie du champ cohérent dans la cavité. Il dépend de la
qualité de la cavité et varie de $\approx$ 150$\mu$s à 1 ms (sans ou avec
anneau réfléchissant). 

T$_d$ temps de décohérence pour le passage de l'état de superposition des
deux états du champ à l'un ou l'autre de ces états. On évalue la
``distance'' {\em D} des deux états du champ à :
$$ D=2\sqrt{n}sin\phi$$
où n est le nombre moyen de photons et $\phi$ l'angle entre les deux
composantes. On a alors :
$$ T_d= 2T_r/D^2 $$

Le temps de décohérence est d'autant plus petit que {\em D} est grand.

Avec n $\approx$ 3,3, T$_r \approx 160 \mu$s, $T_d$ peut descendre à 0,25
T$_r$ rendant possible l'observation d'une manifestation de cette décohérence.

T$_T$ le temps de transit de l'atome est typiquement $\approx$ 100 $\mu$s

Le temps de décohérence t$_D$ pour les détecteurs macroscopiques D est très
petit, trop pour être accessible à la mesure.

\end{quote}

Avec un système mésoscopique on peut comme attendu porter les temps de
décohérence dans le champ des valeurs mesurables et permettre donc
l'observation et l'épreuve quantitative de cette décohérence. Ces temps de
décohérence sont donc largement supérieurs à ceux des détecteurs
macroscopiques D, mais ils restent inférieurs au temps de transit T$_T$ et
c'est donc la décohérence du champ qui a lieu bien avant que l'atome
n'atteigne les détecteurs -au contraire de ce qui a été supposé
jusqu'ici.

Cependant, quels que soient tous ces temps, aucune mesure simple ne permet
de distinguer si l'atome a été mesuré avant le champ ou l'inverse : même la
corrélation des observations de l'état du champ (s'il était observable) et
de celui de l'atome par l'état des détecteurs D reste la même. Il faut une
expérience plus complexe.

\subsection{La décohérence en expérience.}

Pour mettre à l'épreuve ce temps de décohérence après le passage de l'atome
dans la cavité, on va envoyer un second
atome, préparé comme le premier. Selon le temps qui sépare les deux
passages, ce second atome va trouver ou bien le champ dans son état de
superposition, ou bien l'un ou l'autre des états.

La détection des deux atomes de chaque couple peut être trouvée selon quatre
configurations : ee, eg, ge, et gg, et c'est la différence des deux
probabilités conditionnelles :

$$ \eta= \frac{P_{ee}}{P_{ee}+P_{eg}}-
	\frac{P_{ge}}{P_{ge}+P_{gg}} $$

\noindent qui constitue le paramètre pertinent à soumettre à des mesures.

Afin d'assurer l'existence des interférences entre les différentes
``trajectoires'' le mélange quantique des états (celui qui fait passer de e
à e+g) doit être présent {\bf non seulement à l'entrée de la cavité comme
supposé jusqu'ici, mais également à la sortie}. En fait, deux systèmes
identiques alimentés par une même source micro-onde effectuent ces
fonctions.

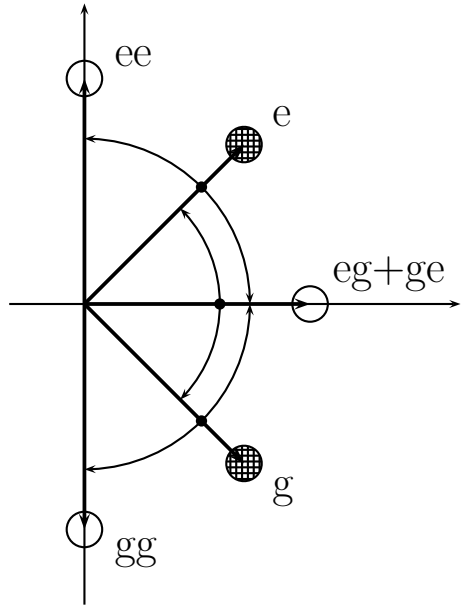
\begin{multicols}{2}

\begin{pspicture}(4,9)
\rput(1.5,4.){

\psline{->}(-1,0)(5,0)
\psline{->}(0,-4)(0,4)
\pscircle*(1.556,1.556){.075}
\pscircle*(1.556,-1.556){.075}
\psarc{<-}(0,0){2.2}{-90}{-45}
\psarc{->}(0,0){2.2}{45}{90}
\psarc{<-}(0,0){2.2}{0}{45}
\psarc{->}(0,0){2.2}{-45}{0}
\pscircle*(1.8,0){.075}
\psarc{->}(0,0){1.8}{0}{45}
\psarc{<-}(0,0){1.8}{-45}{0}
\rput[l]{0}(.4,3.3) {\LARGE {ee}}
\rput[l]{0}(2.5,2.5) {\LARGE {e}}
\rput[l]{0}(3.3,.4){\LARGE {eg+ge}}
\rput[l]{0}(.4,-3.3) {\LARGE {gg}}
\rput[l]{0}(2.5,-2.5) {\LARGE {g}}

\rput[3,3]{-45}(0,0){
\psline[linewidth=.05]{->}(0,0)(3,0)
\pscircle[fillstyle=crosshatch,hatchsep=.05](3,0){.25}
}

\rput[3,3]{0}(0,0){
\psline[linewidth=.05]{->}(0,0)(3,0)
\pscircle(3,0){.25}
}

\rput[3,3]{45}(0,0){
\psline[linewidth=.05]{->}(0,0)(3,0)
\pscircle[fillstyle=crosshatch,hatchsep=.05](3,0){.25}
}

\rput[3,3]{90}(0,0){
\psline[linewidth=.05]{->}(0,0)(3,0)
\pscircle(3,0){.25}
}

\rput[3,3]{-90}(0,0){
\psline[linewidth=.05]{->}(0,0)(3,0)
\pscircle(3,0){.25}
}
}
\end{pspicture}

$~$
\vspace{5mm}

Comme indiqué dans le schéma ci-contre, c'est parce qu'il y a deux chemins
possibles eg et ge qui interfèrent pour revenir finalement à l'état initial
du champ (déphasage zéro) que ce paramètre est sensible à la présence des
états superposés. Le paramètre $\eta$ prend la valeur 0,5 si le champ dans
la cavité est resté dans un état de superposition (un cas pur) quand le
deuxième atome y arrive ; par contre il s'annule si le champ est un mélange.

Mais qu'est-ce qui peut provoquer le pasage de l'un à l'autre? C'est la
mesure (en quelque sorte) de l'état de ce champ par son environnement ; la
décohérence provoquée ici par la fuite des photons hors de la cavité peut
prendre cette fois un temps observable grâce à l'état mésoscopique du champ.
\end{multicols}
En jouant sur la séparation des deux états du champ (déphasage plus ou moins
grand) et sur le taux de fuite (déterminé par le nombre moyen de photons
présents) on rend cette décohérence plus ou moins rapide. Le deuxième atome
trouve un état de superposition du champ ou bien l'un ou l'autre de ces
états du champ.
La valeur moyenne de
$\eta$ explorée en fonction du rapport des temps de décohérence à l'écart de
temps entre les deux atomes passe de 0,5 à 0 selon une loi que l'expérience
semble bien vérifier \cite{Brun96}\cite{Raim01}. On peut considérer que le
concept de décohérence et son traitement théorique quantitatif sont bien
établis expérimentalement.

On va devoir plus bas établir son rapport à l'alternative EPR mais insistons
cependant sur le déroulement global de l'expérience qui consiste en la
répétition d'expériences identiques. Pour chacune de ces expériences,
l'envoi successif de deux atomes séparés par un intervalle de temps donné,
le résultat observé est la séquence de déclenchement des détecteurs e et g
trouvée parmi les quatre possibilités ee eg ge gg. Le paramètre $\eta$ n'est
obtenu qu'après la répétition de nombreuses expériences individuelles. Rien
ne peut être dit de ce qu'était l'état du champ dans chacune des expériences
individuelles. On n'observe la décohérence progressive qu'autant qu'on
observe la décroissance progressive d'un noyau radioactif. Dans les deux
cas, {\bf c'est la probabilité qui évolue progressivement}. Dans les deux
cas le système est dans une alternative d'états (initial/avec particule
émise ; ou de superposition/simple).

\section{Les développements récents et l'alternative EPR.}
\subsection{ La question du
temps.}

Pour la MQ en général et pour la décohérence, on a indiqué plus haut que le
temps n'apparaissait qu'en termes de séquences de différences de temps
séparant la réalisation d'événements successifs. Le positionnement absolu de
cette séquence est non seulement inconnu -c'est assez évident- mais non
maîtrisable. Si l'observation et l'intervention ultime de la conscience
jouent un rôle dans la mesure, leur positionnement, au moins précis, dans les
séquences temporelles semble délicat à établir. Il faut pourtant qu'elles
terminent ces séquences. Examinons comment ce temps ``macroscopique'' est
abordé par les différents intervenants.

Pour EPR, il y a d'une part un temps au-delà duquel l'interaction entre les
deux systèmes est terminée, d'autre part un laps de temps ultérieur où un
dispositif de mesure ou un autre est mis en place, et enfin un temps de
mesure avec le dispositif choisi. C'est précisément la constatation d'une
dépendance du résultat du choix du dispositif {\em après} que l'interaction
ait été déclarée terminée qui fait problème. Pour EPR, ``dépendre de'' est
identique à ``interagir avec''.
Si on est amené à parler d'interaction après l'interaction,
c'est que l'interaction n'est pas terminée quand on prétend qu'elle l'est.
Rappelons que l'autre terme de l'alternative EPR -la dépendance est issue
du passé commun- n'est pas défendable aujourd'hui, on verra plus loin qu'on
peut douter qu'elle l'ait encore été pour Einstein en 1949.

\newpage
Si Bohr n'a pas traité ce terme de l'alternative EPR, il a cependant évoqué
la question du temps :
\bq
The experimental arrangements hitherto discussed present a special
simplicity on account of the secondary role which the idea of time plays in
the description of the phenomena in question. It is true that we have freely
made use of such words as ``before'' and  ``after'' implying time
relationships ; but in each case allowance must be made for a certain
inaccuracy, which is of no importance, however, so long as the time
intervals concerned are sufficiently large compared with the phenomenon
under investigation {\em \cite{Bohr35} page 700.}
\eq

Pour Schrödinger en complément de ce que nous avons déjà rapporté,
d'un côté, une position très claire :
\bq
Since they  [Q.M. predictions] are supposed to relate measurement results,
they would be entirely without content if the relevant variables were not
measurable {\em for} a definite point of time, whether the measurement itself
lasts a long or a short while.

When we {\em learn} the result is of course quite immaterial {\em
\cite{Schr35} page 166.}.
\eq
et de l'autre une déclaration qui devient plus embarrassée quand on entre
dans le détail (on reprend et poursuit la citation du chapitre 10, page 162)
:
\bq
\ldots what befalls it is already past.
But it would not be quite right to say
that the $\psi$-function of the object which changes {\em otherwise}
according to a partial differential equation, independent of the observer,
should {\em now} change leap fashion because of a mental act. For it had
disappeared ; it was no more. Whatever is not, no more can it change. It is
born anew, is reconstituted, is separated out from the entangled knowledge
that one has, through an act of perception, which as a matter of fact is
not a physical effect on the measured object. From the form in which the
$\psi$-function was last known, to the new in which it reappears, runs no
continuous road -it ran indeed through annihilation. Contrasting the two
forms, the things look like a leap. In truth something of importance happens
in between, namely the influence of the two bodies on each other, during
which the object possessed no private expectation-catalog nor had any claim
thereto, because it was not independent {\em \cite{Schr35} page 162.}
\eq

On observe là un mélange peu convaincant entre  le recours à
l'action mentale déjà apparu et un rappel opportun de la physique mais qui
ne survit pas pour autant.

\subsection{Ce que la décohérence introduit dans le débat. }

C'est dans le formalisme de la matrice densité, que la décohérence prédit le
passage d'un cas pur à un mélange par l'extinction progressive des termes
non-diagonaux/d'interférence. Grâce à ses termes diagonaux, la décohérence
prédit aussi le comportement statistique des valeurs observées dans l'examen
d'un ensemble. Mais {\bf elle ne dit rien sur l'émergence pour une
expérience particulière de l'un ou l'autre des cas possibles, et seulement
l'un}. On peut même dire que le formalisme de la matrice densité est bien
adapté pour faire disparaître cette carence. En ce sens elle ne répond pas
non plus au questionnement d'Einstein concernant l'aspect probabiliste de la
MQ.

Mais comment la décohérence répond-elle à l'alternative EPR ? et comment
permet-elle d'éviter
le recours à la conscience devenu aujourd'hui encombrant?

Présenté dans les termes objectifs du réalisme il est vrai, la décohérence
introduit une constante de temps entre la mise en rapport du dispositif
de mesure et du système mesuré, et la séparation
irréversible (et plus tard observable) de l'une des réalisations de couples
de variables intriquées (on notera cependant une certaine imprécision quant
au temps zéro, il s'agit le plus souvent de corrélations, de séquences
temporelles).
Sans y obliger, la décohérence pousse ainsi vers la description de la
``mesure'' comme un phénomène physique concret, objectif. Ce que justement
la ``volte-face épistémologique'' avait voulu éviter avec son recours à la
conscience. Avec la décohérence, il semblerait alors que ce recours à la
conscience soit devenu plus nécessaire. Mais ce recours déjà en principe
indéfendable apparaît aujourd'hui de moins en moins acceptable dans la
communauté des physiciens.

On comprend peut-être alors -on comprend sans justifier!- le contresens sur
le chat de Schrödinger et le questionnement sur le passage  du micro- au
macroscopique :
\bq
This result illustrates the basic feature of the quantum to classical
transition. Mesoscopic superpositions made of a few quanta are expected to
decohere in a finite time interval [shorter than $T_r$], while macroscopic
ones ($n\gg$1) decohere instantaneously and cannot be observed in
practice {\em \cite{Brun96} page 4887}.
\eq
Si le problème avait été celui là -la non-observation directe des états de
superposition- alors il serait résolu et l'interprétation de Copenhague
émergerait immaculée. Il n'en est rien.

En parallèle avec cet usage de la décohérence et bien antérieurement, on a
parlé de non-localité, de non-séparabilité (des expressions forgées pour cet
usage), mais si ces expressions peuvent être utiles comme raccourci de
description, elles doivent appeler les questions sur le pourquoi et le
comment de ces propriétés, elles ne peuvent les remplacer. On ne peut
accepter un comportement des physiciens vis-à-vis de la microphysique
comparable à celui des médecins 
de Molière vis-à-vis de
l'opium (Le Malade imaginaire, Intermède 3):
\bq
\begin{center}
\begin{verse}
\begin{tabular}{ll}
\multicolumn{1}{c}{BACHELIERUS} &
\multicolumn{1}{c}{LE BACHELIER}\\
 Mihi a docto doctore & Par le docte docteur il m'est \\

  Demandatur causam et
rationem quare & Demandé la cause et la raison pour lesquelles\\

Opium facit dormire. & L'opium fait dormir.\\

A quoi respondeo, & A quoi je réponds :\\

Quia est in eo & Parce qu'il est en lui\\

Vertus dormitiva, & Une vertu dormitive,\\

Cujus est natura & Dont la nature est \\

Sensus assoupire & d'assoupir les sens. \\

\multicolumn{1}{c}{CHORUS} &
\multicolumn{1}{c}{LE CH\OE UR} \\
Bene, bene, bene, bene respondere & Bien, bien, bien, bien répondu :\\

Dignus, dignus est intrare & Digne, il est digne d'entrer\\

In nostro docto corpore & Dans notre docte corporation.\\

Bene, bene, bene, bene respondere & Bien, bien, bien, bien répondu :\\

\end{tabular}

\end{verse}

\end{center}
\eq

\vspace{6mm}

Alors, une vertu non-locale? Dont la nature est d'étendre la portée?

Rappelons-en le contenu dans les termes d'Einstein (EPR) :

\bq On this point of view since either one or the other, but not both
simultaneously, of the quantities P and Q can be predicted, they are not
simultaneously real.
This makes the reality of P or Q [sur le système II] depend upon the process
of measurement carried out on the first system which does not disturb the
second system in any way. No reasonable definition of reality could be
expected to permit this.
\eq

C'est la contradiction entre ``depend upon'' et ``does not disturb'' qu'il
faut lever : il y a dépendance et interaction dans des termes spécifiques
qu'il faut préciser. Il faut aussi préciser l'articulation de cette
dépendance ou de cette interaction avec la relativité restreinte sans se
contenter de dire ``{\em  There is never contradiction between nonlocality and
relativistic causality'' {\em \cite{Raim01} page 566.}}

\subsection{Retour à Einstein et retour d'Einstein.}
Examinons ce que sont devenus en 1949 les trois éléments qui en 1926
caractérisaient pour Albert Einstein la MQ selon sa 
lettre à M. Born (citée plus haut) :  le
respect pour les succès de la MQ, l'insatisfaction globale, et
l'insatisfaction plus nette
concernant l'aspect probabiliste des résultats. Il commence par
ce dernier qui reste il faut le dire le fer de lance :
\bq
I am, in fact, [au contraire ``of almost all contemporary theoritical
physicists''] firmly convinced that the essentially statistical character of
contemporary quantum theory is solely to be ascribed to the fact that this
[theory] operates with an incomplete description of physical
systems {\em \cite{Eins49} page 666}.
\eq
Notons qu'aucun recours, direct ou non, à d'éventuels paramètres cachés n'est
ici évoqué. Il poursuit par le respect aux succès de la MQ qu'il précise
avec la plus grande clarté :
\bq
This theory [statistical quantum theory] is until now the only one which
unites the corpuscular and ondulatory dual character of matter in a
logically satisfactory fashion ; and the (testable) relations which are
contained in it, are, within the natural limits fixed by the indeterminacy
relation, {\em complete.} The formal relations which are given in this
theory -i.e. its entire mathematical formalism- will probably have to be
contained, in the form of logical inferences, in evey useful future theory
 {\em \cite{Eins49} pages 666-667}.
\eq
Un éloge impressionnant qui contredit radicalement l'opinion quelquefois
attribuée à Einstein d'un rejet de la MQ.
Il passe ensuite au ``paradoxe'' EPR :
\bq
By this way of looking at the matter [la manière de Bohr] : ``the partial
systems A and B form a total system''] it becomes evident that the paradox
forces us to relinquish one of the following two assertions :
\bq
(1) the description by means of the $\psi$-function is {\em complete}

(2) the real states of spatially separated objects are independent of each
other.
\eq
On the other hand, it is possible to adhere to (2), if one regards the
$\psi$-function as the description of a (statistical) ensemble of systems
(and therefore relinquishes (1)). Howecer, this view blasts the framework of
the ``orthodox quantum theory.'' \cite{Eins49}{\em page 682}.
\eq

On retrouve une alternative plutôt qu'un paradoxe, mais chacun des
termes conduit bien à une incomplétude de la MQ. Notons que l'abandon de
(2) ne donne aucune réponse à l'origine du caractère probabiliste du
résultat des mesures, mais on admet la non-séparabilité.

Que ce soit là pour Einstein une question de physique nous paraît attesté par ce
commentaire final -insistons qu'il termine toute la discussion concernant la
MQ par cela- :

\bq
I close these expositions, which have grown rather lengthy, concerning the
interpretation of quantum theory with the reproduction of a brief
conversation which I had with an important theoretical physicist.

He : ``I am inclined to believe in telepathy.''

I : `` This has probably more to do with physics than with psychology.''

He : ``Yes''

{\em \cite{Eins49} page 683}.
\eq

Cela devrait interpeller étrangement les auteurs des commentaires que l'on peut
lire aujourd'hui :

\bq
According to our calculations, we may decide whether to emphasize wave like
(interference) or particle like (which path) behavior even after the emission
is over {\bf without physically ``manipulating''} the $\gamma$
photons \cite{Scul82}{\em Marlan O. Scully 1982, page 2208}.
\eq

\bq
Pourtant il {\em [le photon]} se comporte toujours d'une façon qui dépend
du test effectué sur l'autre photon, bien {\bf qu'il ne puisse être physiquement
influencé} par l'accomplissement de la mesure ou par le résultat ainsi
obtenu {\em \cite{Ghir02}
 Giancarlo Ghirardi 2002, page 22.}
\eq

\bq
On reviendra sur cette question\ldots pour décrire comment l'expérience
[EPR] a été réalisée en soulignant ce qui semble faire aujourd'hui son
intérêt essentiel : la mise en évidence d'états ``enchevêtrés'' à longue
distance \ldots avec toutes leurs {\bf diableries} surprenantes mais
avérées {\em \cite{Omne00} R. Omnès 2000, page 168.}
\eq

\bq
Cette intriquation subsiste même si les deux atomes se sont éloignés l'un de
l'autre et se trouvent séparés après la collision par une distance
arbitrairement grande. Elle décrit une ``non-localité'' fondamentale de la
physique quantique. Une mesure de l'atome 1 peut avoir un effet immédiat à
grande distance sur le résultat de la mesure de l'atome 2! Il y a donc entre
les deux particules un lien quantique {\bf immatériel et instantané}. C'est
Einstein, avec ses collaborateurs Podolsky et Rosen, qui a discuté le
premier en 1935 cet aspect troublant de la théorie quantique {\em
\cite{Haro01} Serge Haroche (2001) page 577.}
\eq

Mais justement, peut-être Einstein répondrait-il à tous ces physiciens que
c'est de physique qu'il s'agit et d'une part que l'immatériel n'a pas sa
place et que d'autre part, l'instantané pose problème.

Le nouveau cependant, c'est bien comme le déclarent les 
intervenants de l'ENS :
\bq
On peut alors aborder à nouveau, mais de façon concrète, l'étude des
fondements de la théorie.{\em Voir par exemple \cite{Haro01} page 572.}
\eq
Oui, les questions posées il y a cinquante ans subsistent. On peut bien les
définir comme fondements de la théorie, ce qu'ils sont aussi, mais ce sont
comme le répétait Einstein les marges incertaines, insatisfaisantes,
incomplètes, inachevées, de cette théorie. Elles sont devenues accessibles à
l'exploration : les distances sont macroscopiques (ici quelques centimètres)
et 
les temps aussi (durées de l'ordre de la milliseconde).
Un programme d'exploration
existe, mais pourquoi ne pas l'infléchir pour interroger directement cette
``vertu non-locale''?
\section{Conclusion.}

Nous avons montré que le recours fréquent à l'expérience de pensée dite ``du
chat de Schrödinger'' était utilisé à contresens. Cela restera anecdotique
si ce contresens n'est pas utilisé pour fuir les questions posées en 1935
par Einstein, Podolsky et Rosen.

L'essentiel est en effet dans l'argument /l'alternative /la démonstration
EPR et dans la réponse d'abord proposée par Schrödinger et reprise pendant
60 ans d'un recours direct ou indirect à un rôle pour la conscience humaine
dans cette question de physique, réponse devenue aujourd'hui encombrante.

L'argument EPR et les développements ultérieurs d'Einstein {\em montrent}
que la MQ est tout aussi radicalement correcte qu'incomplète ou inachevée.

La MQ prévoit en effet deux types d'évolution sans être réellement capable
de préciser quand l'un ou l'autre doivent être utilisés. Par ailleurs, elle
ne s'avère capable de prévoir, même en principe, que le devenir d'un
ensemble de systèmes préparés identiquement mais pas celui d'un système
individuel particulier.

Enfin, dans le processus de mesure lui-même, qui n'est qu'une réalisation
particulière de l'interaction temporaire de deux sous-systèmes initialement
séparés, elle prévoit des interactions à distance, corrélations instantanées
à distance, baptisées effets ``non-locaux'', simplement constatés mais ni
compris ni explicités.

La ``volte-face épistémologique'' introduite par Schrödinger et pratiquée
depuis avec plus ou moins de clarté ou d'hypocrisie, consiste en l'abandon
du réalisme. Elle conduit à donner un statut de catalogue à la fonction d'onde
et un rôle spécifique essentiel à l'observateur humain et à sa conscience.

Que peut-on dire de la situation actuelle?

Notons pour mémoire (le point n'a pas été abordé dans le texte) que
l'introduction de variables/paramètres supplémentaires par L. De Broglie, D.
Bohm et J.P. Vigier \cite{Brog52} permet de construire une MQ déterministe 
(l'aspect probabiliste est ainsi expliqué), mais c'est au prix d'un recours
accru à une non-localité qui reste tout autant essentielle, arbitraire et
inexpliquée.

On se doit de redire avec force que  la volte-face épistémologique avec son
encombrant recours à la conscience ne peuvent être évités que par un retour
au réalisme qui porte alors en pleine lumière la manifestation d'effets à
distance parce qu'ils sont alors entièrement de l'ordre de la physique.

L'élaboration du concept de décohérence qui modèlise introduit/traite
l'interaction d'un système microscopique avec son environnement peut sembler
amorcer un retour au réalisme : il décrit un processus avec une durée
déterminée. Sans chercher à mettre en question la valeur de sa
description, on peut noter qu'il explique bien le passage d'un cas pur à un
mélange (dans certaines conditions, il prédit ainsi une durée de vie
incroyablement petite pour un état de superpodition chat mort/ chat
vivant!). Il s'avère par contre incapable d'expliquer la venue de l'une ou
l'autre des réalisations dont il prédit pourtant bien la probabilité
correspondante/attendue. Le problème de la mesure reste alors entier.

Mais quand bien même la décohérence arriverait à la description de la venue
d'une seule des réalisations possibles, par exemple au travers de paramètres
supplémentaires (cette fois bien cachés!) dans l'environnement, elle
aboutirait alors elle aussi à mettre en lumière la même réalité des effets
instantanés à distance : la non-localité, qui reste bien aujourd'hui non
seulement un élément de réalité physique incontournable mais justement et à
cause de cela un {\bf objet d'étude en soi.}

Les développpements expérimentaux et théoriques qui s'articulent autour de
l'optique quantique, les manipulations d'atomes et de champ cohérents dans
une cavité ont déjà permis d'accéder à des expériences mettant en jeu ces
effets ; elles peuvent plus directement prendre ces effets comme {\bf objet
d'étude}.

Il reste indispensable de garder un contact direct avec les {\bf textes
fondateurs}. Cela évite d'abord de reproduire des erreurs dans leur
utilisation, c'est la moindre des choses. Mais surtout, leur rigueur et leur
richesse permet d'appréhender plus en profondeur les processus à l'\oe uvre
dans les réalisations expérimentales et théoriques actuelles.

\end{document}